\documentclass[draft]{agujournal2019}
\def\degree{\hbox{$^\circ$}}
\usepackage{url} 
\usepackage[inline]{trackchanges} 
\usepackage{soul}

\usepackage{amsmath}
\usepackage{amssymb}

\draftfalse

\journalname{AGU Advances}

\begin{document}

%
%


\title{Non-synchronous rotation on Europa driven by ocean currents}

%
%




\authors{Yosef Ashkenazy\affil{1}, Eli Tziperman\affil{2}, Francis Nimmo\affil{3}}

\affiliation{1}{Department of Solar Energy and Environmental Physics, The Blaustein Institutes for Desert Research, Ben-Gurion University of the Negev, Midreshet Ben-Gurion, 84990, Israel.}
\affiliation{2}{Department of Earth and Planetary Sciences and School of Engineering and Applied Sciences, Harvard University, Cambridge, MA, 02138, USA}
\affiliation{3}{Department of Earth and Planetary Sciences, University of California Santa Cruz, Santa Cruz, CA 95064, USA}





\correspondingauthor{Yosef Ashkenazy}{ashkena@bgu.ac.il}




\begin{keypoints}
\item Ocean currents on icy moons such as Europa may drive rotation of their overlying ice shells
\item A mathematical model for ice shell movements driven by ocean stress, shell elasticity and viscous adjustment is developed
\item The model shows rotation rates slightly slower than synchronous and is used to constrain ice shell parameters such as effective viscosity
\end{keypoints}

%
%

%
%


\begin{abstract}
  It has been suggested that the ice shell of Jupiter's moon Europa may drift non-synchronously due to tidal torques. Here we argue that torques applied by the underlying ocean are also important and can result in non-synchronous rotation (NSR). The resulting spin rate can be slightly slower than the synchronous angular rate that would have kept the same point of the ice shell facing Jupiter. We develop an ice shell rotation model, driven by ocean stress calculated using a high-resolution state-of-the-art ocean general circulation model, and take into account the viscoelastic deformation of the ice shell. We use the ice shell model results together with observed limits on the ice shell drift speed to constrain ice shell parameters such as effective viscosity, which is currently uncertain by at least four orders of magnitude. Our results suggest, at best, sluggish ice shell convection. Depending on the relaxation time scale of the ice shell and on the ocean currents, the ice shell may exhibit negligible drift, constant drift, or oscillatory drift superimposed on random fluctuations. The expected rotation rate exceeds $\sim$30~m/yr; future spacecraft observations can be used to test these predictions and yield insight into the properties of the ice shell and underlying ocean.
\end{abstract}

\section*{Plain Language Summary}

Some icy moons in the solar system, like Jupiter's moon Europa and Saturn's moon Enceladus, are believed to have deep oceans below their ice shell. Such moons, Europa included, may also be tidally locked to their corresponding planet such that the same side of the moon always faces its planet. State-of-the-art oceanic simulations of Europa's ocean exhibit strong upper ocean jets. We propose that these jets can drive its overlying ice shell, to drift slowly from its tidally locked, synchronized rotation state. We develop a mathematical model to study the effects of ocean currents on the overlying ice shell. We show that the ocean currents may cause the ice shell to drift and that future measurements of the drift may be used to estimate key unknown parameters of the ice shell such as its effective viscosity.  

%
%

\section{Introduction}
\label{sec:introduction}
Europa, one of the four Galilean moons of Jupiter, is considered a prime candidate for extra-terrestrial life \cite{Chyba-Phillips-2001:possible, Hand-Chyba-Priscu-et-al-2009:astrobiology,Pappalardo-Vance-Bagenal-et-al-2013:science} due to its deep ($\sim$100 km) ocean \cite{Cassen-Reynolds-Peale-1979:is, Carr-Belton-Chapman-et-al-1998:evidence, Kivelson-Khurana-Russell-et-al-2000:galileo} that underlies a thick ice shell (several to tens of km) \cite{Billings-Kattenhorn-2005:great,Cassen-Reynolds-Peale-1979:is, Carr-Belton-Chapman-et-al-1998:evidence, Hussmann-Spohn-Wieczerkowski-2002:thermal, Tobie-Choblet-Sotin-2003:tidally}.  The existence of an ocean under the ice shell is indicated by the observed induced magnetic field \cite{Khurana-Kivelson-Stevenson-et-al-1998:induced}, observations of ice tectonics \cite{Pappalardo-Belton-Breneman-et-al-1999:does} and perhaps also by water vapor plumes over Europa's mid-southern latitudes \cite{Roth-Saur-Retherford-et-al-2014:transient, Sparks-Hand-McGrath-et-al-2016:probing}.

Europa is tidally locked to Jupiter such that its spin rate (Europa's day is equal to about 3.5 Earth days) is approximately equal to the orbital rotation rate around Jupiter. The maximum possible non-synchronous rotation (NSR) rate was estimated to be $\sim$1 km~yr$^{-1}$ \cite{Hoppa-Greenberg-Geissler-et-al-1999:rotation} based on a comparison between Voyager 2 and Galileo spacecraft images that are 17 years apart. There are several hints that Europa rotates non-synchronously. First, the distribution of craters on Europa does not seem to reflect the asymmetry one would expect under complete synchronization, as the leading hemisphere should host more craters in comparison to the trailing hemisphere \cite{Shoemaker-Wolfe-1982:cratering, Chapman-Merline-Bierhaus-et-al-1998:cratering}. Second, mapping the fine-scale fractures and patterns of Europa's surface {has been used to argue for} non-synchronous rotation \cite{Helfenstein-Parmentier-1985:patterns, McEwen-1986:tidal, Geissler-Greenberg-Hoppa-et-al-1998:evidence, Greenberg-Geissler-Hoppa-et-al-1998:tectonic}. 

Non-synchronous rotation can occur either for the entire moon or for the ice shell separate from the interior \cite{Greenberg-Weidenschilling-1984:how, Hoppa-Greenberg-Geissler-et-al-1999:rotation}. It can be driven by tidal torques associated with the eccentric orbit of Europa that can slightly increase its rotation rate \cite{Hoppa-Greenberg-Geissler-et-al-1999:rotation}, perhaps combined with slow thermal adjustment of the ice shell \cite{Ojakangas-Stevenson-1989:thermal}. However, a sufficiently large mass or shape asymmetry, possibly together with an elastic resistance of its ice shell, can lead to complete phase locking (synchronous rotation) \cite{Greenberg-Weidenschilling-1984:how, Hoppa-Greenberg-Geissler-et-al-1999:rotation,Goldreich-Mitchell-2010:elastic}. If Europa's ocean decouples the rocky core from the ice shell, the shell can display non-synchronous rotation even when the rocky core of Europa is phase locked \cite{Hoppa-Greenberg-Geissler-et-al-1999:rotation}. Below we focus on the possibility of non-synchronous rotation of the ice shell driven by ocean currents, {which has not been investigated in a quantitative fashion hitherto}.

Europa's ocean dynamics have been studied using a variety of models and mechanisms \cite{Thomson-Delaney-2001:evidence, Goodman-Collins-Marshall-et-al-2004:hydrothermal, Melosh-Ekholm-Showman-et-al-2004:temperature, Tyler-2008:strong, Vance-Goodman-2009:oceanography, Goodman-2012:tilted, Goodman-Lenferink-2012:numerical, Soderlund-Schmidt-Wicht-et-al-2014:ocean, Gissinger-Petitdemange-2019:magnetically}, and scaling arguments were used to suggest the existence of alternating zonal jets \cite{Vance-Goodman-2009:oceanography}. Tides can also excite internal waves \cite{Rovira-Navarro-Rieutord-Gerkema-et-al-2019:do} and libration-driven elliptical instability can also drives ocean motions \cite{Lemasquerier-Grannan-Vidal-et-al-2017:libration}. Recent studies of Europa's ocean \cite{Soderlund-Schmidt-Wicht-et-al-2014:ocean, Soderlund-2019:ocean, Ashkenazy-Tziperman-2021:dynamic, Kang-Mittal-Bire-et-al-2022:how, Kang-2022:different, Zeng-Jansen-2021:ocean} used global models, taking into account elements such as non-hydrostatic effects and the full Coriolis force, to study the ocean dynamics, and reported a wide low-latitude eastward jet, high-latitude westward jets, and a highly turbulent ocean. {Negative (westward), upper ocean, low latitude zonal flows have been reported by some previous studies \cite{Soderlund-Schmidt-Wicht-et-al-2014:ocean, Ashkenazy-Tziperman-2021:dynamic, Kang-2022:different} and can be attributed to the thermal-wind relation in which the zonal velocity decreases with height if the density decreases poleward; see  \citeA{Ashkenazy-Tziperman-2021:dynamic}.}

The zonal jets at the top of the ocean exert stress on the bottom of the ice shell, which can cause a slow drift of the shell relative to the rocky core. In the present study, we develop a model of ice shell drift that takes into account the influence of oceanic stress on the dynamics of the ice shell, as well as the viscoelastic adjustment of the ice shell itself. We ignore internal ice flow within the shell as this is very slow in comparison to the ice shell drift discussed here \cite{Ashkenazy-Sayag-Tziperman-2018:dynamics}. By comparing the model predictions to observational constraints, we estimate and constrain parameters of the elastic and viscous responses of the ice shell that are currently very poorly known.

To understand the interaction of the non-synchronous ice shell drift and the viscoelastic response of the shell, we need to consider both the location of the tidal bulge (the direction of the long axis of its ellipsoidal shape) and of some feature (e.g., a crater) on the ice shell surface assumed to initially face Jupiter. First, we consider two limit cases of possible movement of the ice shell in response to an ocean torque: (a) The ice shell is rotated as a rigid body without any deformation of the ice. In that case (the rigid shell case), tidal torques will act to restore the ice shell bulge to face Jupiter, and no elastic restoring force will be active. (b) The ice shell tidal bulge remains facing Jupiter, yet the ice shell itself rotates and deforms such that the location of a crater propagates away from the line connecting the centers of Europa and Jupiter (the flexible shell case). In this case, the net torque due to the tidal forces on the ice shell vanishes (because it acts on the bulge which still faces Jupiter), and only the force due to the ice shell elasticity \cite{Goldreich-Mitchell-2010:elastic} will act to restore the crater to its original location. In reality, both forces (tidal and elastic) act on the ice shell. Since the tidal force is much larger than the elastic force \cite{Goldreich-Mitchell-2010:elastic}, the tidal bulge of the ice shell remains facing Jupiter. However, as long as the shell is sufficiently flexible, it can rotate under the influence of ocean torques while conforming its shape to the tidal bulge that remains facing Jupiter. We thus assume below that the tidal bulge faces Jupiter and ignore the tidal force in our subsequent calculations.

In the above, we did not consider the viscous effects of the ice, which may affect the non-synchronous rotation of the ice shell as follows. In the absence of viscous effects, if the ice shell is initially rotated by an ocean torque, elastic torques will attempt to return the crater to its original position, and the shape will be restored such that the tidal bulge is back at the crater's position. However, viscous ice adjustment means that the bulge relaxes to its new position relative to the crater \cite{Greenberg-Weidenschilling-1984:how}, which weakens the elastic force with time, such that the deformed ice will not return to its original shape. The ocean torque can now lead to a further motion of the crater and again to a viscous adjustment to the new position. This amounts to a slow drift of the crater location in response to a continuous torque due to ocean currents. The speed of the drift depends on the viscous adjustment/relaxation time scale of the ice \cite{Greenberg-Weidenschilling-1984:how}---faster drift when the viscous time scale is short.



%
%
%
%

\section{Methods and Model}

\subsection{Ocean simulations and the calculation of the ocean torque}
\label{sec:ogcm}

The ocean simulations are two-dimensional (latitude-depth) and were performed using the MITgcm, a state-of-the-art oceanic general circulation model \cite{Marshall-Adcroft-Hill-et-al-1997:finite, MITgcm-manual-github:mitgcm}. We use no-slip boundary conditions at the bottom and at the ocean-ice interface, with and without a linear drag term applied at the ocean top and bottom levels. The meridional resolution is 1/24 of a degree latitude (1.1 km) spanning a meridional range from 70\degree{S} to 70\degree{N}. There are 100 vertical levels with thicknesses from 25~m near the top to 1150~m at the bottom where the overall depth of the ocean is 100 km. We use the shelf ice package of MITgcm \cite{Losch-2008:modeling} to represent a 10 km thick ice shell. The surface temperature is prescribed following \citeA{Ashkenazy-2019:surface}. The eddy coefficients follow the choices of \citeA{Ashkenazy-Tziperman-2021:dynamic} where the horizontal eddy diffusion for temperature and salinity is 30~m$^2$~s$^{-1}$, the horizontal viscosity is 300 m$^2$~s$^{-1}$, the vertical diffusion is $10^{-4}$ m$^2$~s$^{-1}$, and the vertical viscosity is $10^{-3}$ m$^2$~s$^{-1}$. The ocean linear friction coefficient $\tilde{r}_o$ is set to  $2\times10^{-4}$ m~s$^{-1}$, a typical value used in Earth's ocean modeling \cite{Marshall-Adcroft-Hill-et-al-1997:finite, MITgcm-manual-github:mitgcm}; we also used a three times higher linear friction coefficient of $\tilde{r}_o=6\times10^{-4}$ m~s$^{-1}$ to verify the sensitivity of the results to this parameter.

Using the output of the ocean model, we calculate the zonal stress on the ice shell as follows,
\begin{linenomath*}
\begin{align}
  \tau_\lambda(\lambda,\phi)&=-\rho_o \left(\nu_z+\tilde{r}_o  
  \frac{\Delta z}{2}\right)  
  \left.\frac{\partial u}{\partial z} \right\vert_{z=0} \nonumber \\
  &\approx-\rho_o \left(\nu_z+\tilde{r}_o  \frac{\Delta z}{2}\right)
  \frac{R\omega\cos\phi-u({z=-\Delta z/2})}{{\Delta z}/{2}}
  \label{eq:tau_lambda}
\end{align}
\end{linenomath*}
where the second line uses the finite difference approximation of the vertical shear, taking the difference between the surface ocean velocity which is equal to that of the ice shell (no-slip condition), and the ocean velocity in the middle of the uppermost ocean level, $\Delta z/2$. Also, $\lambda$ is the longitude, $\phi$ is the latitude, $\rho_o=1046$ kg~m$^{-3}$ is the ocean water density, $\nu_z$ is the oceanic vertical viscosity coefficient, $\omega$ the angular velocity of the ice shell, and $R=1561$~km is the radius of Europa. The stress $\tau_\lambda$ is used to calculate the torque in the $z$ direction (parallel to the rotation axis) using an integral of the zonal stress $\tau_\lambda$ times the distance from the axis of rotation ($R\cos\phi$), integrated over the surface area within the model domain,
\begin{linenomath*}
\begin{align}
    \frac{M_z}{I}&=\frac {R^3}{I} \int\tau_\lambda \cos^2\phi \;d\phi\;d\lambda 
    =r_o (\omega_o-\omega) \label{eq:Mz}
\end{align}
\end{linenomath*}
where 
\begin{linenomath*}
\begin{align}
    r_o&=\frac{16\pi R^4\rho_o}{3I\Delta z}\left(\nu_z+\tilde{r}_o\frac{\Delta z}{2}\right), \label{eq:r0}
\\
    \omega_o&=\frac{3}{8\pi R}\int u\left({z=-\frac{\Delta z}{2}}\right) \cos^2\phi \;d\phi\;d\lambda.\label{eq:omega0}
\end{align}
\end{linenomath*}
Then the torque is multiplied by an appropriate constant---{$24\times360=8640$ in our case as the model's resolution is 1/24 of a degree}---to represent the torque applied by a 360\degree{} longitude ocean. 
We calculate the moment of inertia of the ice shell as $I=8\pi\rho_i R^5 (1-\alpha^5)/15=4.5\times10^{32}$ kg~m$^2$, where $\alpha=(R-H_{\mathrm{ice}})/R$, the ice shell density is $\rho_i=917$~kg~m$^{-3}$, and $H_{\mathrm{ice}}=10$~km. The uncertainty of the latter value is comparable to the ice thickness itself. Following Eq.~\eqref{eq:r0}, in the absence of linear drag in the ocean, $r_o\sim 9.2\times 10^{-9}$s$^{-1}$ while the value is higher when the oceanic drag coefficient is taken into account; i.e., $r_o\sim 3.2\times 10^{-8}$~s$^{-1}$ when using a standard ocean linear drag of $\tilde{r}_o=2\times 10^{-4}$ m~s$^{-1}$ and $r_o\sim 7.9\times 10^{-8}$~s$^{-1}$ when using a three times larger ocean linear drag of $\tilde{r}_o=6\times 10^{-4}$ m~s$^{-1}$. We note that different linear drag coefficients $\tilde{r}_o$ result in different surface currents and thus different forcing; thus the friction coefficient $r_o$ cannot be altered without altering the ocean forcing $\omega_o$. However, our results show that all simulations with and without oceanic drag $\tilde{r}_o$ yield similar constraints on $\tau$. We note that we performed only three, 2D, oceanic simulations to estimate the torque due to ocean currents, and found that all resulted in similar retrograde drift; by retrograde drift, we are referring to the ice shell spinning slightly slower than the synchronous angular rate that would have kept the same point of the ice shell facing Jupiter. This retrograde flow is consistent with previous studies \cite{Soderlund-Schmidt-Wicht-et-al-2014:ocean, Ashkenazy-Tziperman-2021:dynamic, Kang-2022:different} that reported westward flow at the upper ocean of the low latitudes.

Given the two-dimensional geometry of the model, the torque due to the meridional flow $\tau_\phi$ vanishes due to symmetry. That is, meridional stress (say due to a poleward surface flow in the northern hemisphere) implicitly exists at all longitudes. When this stress is integrated over the ocean surface, the net poleward torque therefore vanishes (the poleward stress at any longitude is canceled by an equal contribution at a longitude 180\degree{} away). Thus the ocean model used here cannot represent meridional ocean torques on the ice shell. 

\subsection{Model for the ice shell drift rate}
\label{sec:icy-shell-model}

\begin{figure}[!tbhp]
    \centering
    \includegraphics[width=0.9\textwidth]{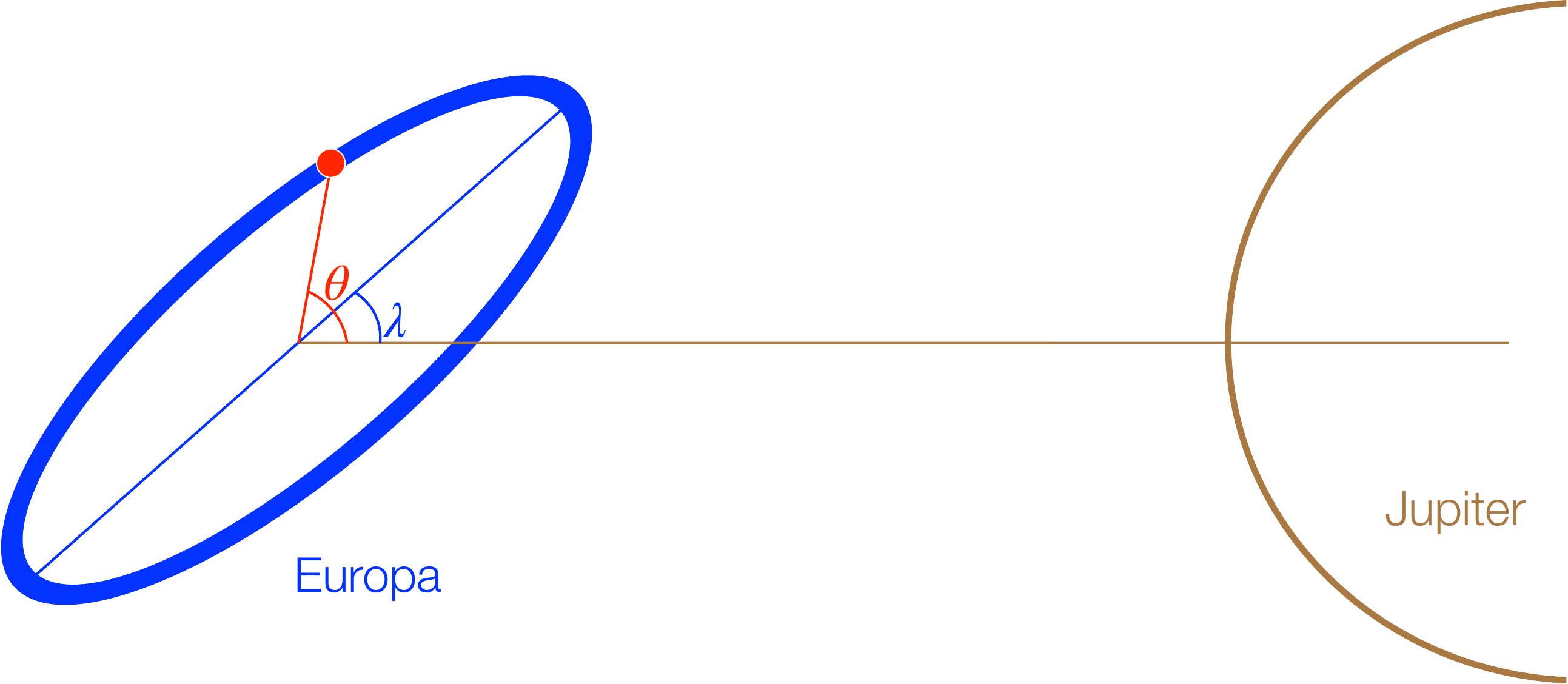}
    \caption{A schematic showing the Europa ice shell on the left and Jupiter on the right, with the angle $\theta$ appearing in the model equations \eqref{eq:model1}. {We assume that $\lambda$=0}.}
    \label{fig:Europa-icy-shell-drift-schematic}
\end{figure}

In the model proposed here, the position of the ice shell is represented by an angle $\theta$ between a fixed location on the equatorial plane of Europa's ice shell (e.g., the location of a crater) and the axis connecting the centers of Jupiter and Europa, see Fig.~\ref{fig:Europa-icy-shell-drift-schematic}. We assume, based on \citeA{Goldreich-Mitchell-2010:elastic}, that the tidal torque is much larger than the elastic torque such that the tidal bulge raised by Jupiter is facing Jupiter with no offset, that is, $\lambda=0$ in Fig.~\ref{fig:Europa-icy-shell-drift-schematic}. In that case, the momentum equation that describes the motion of the ice shell is,
\begin{linenomath*}
\begin{align}
    &\frac{d^2\theta }{dt^2}+F_e
    =\frac{M_z}{I} 
    =r_o\left(\omega_o-\frac{d\theta}{dt}\right)\nonumber  \\
  &F_e=k_i \left[\theta(t)-\frac{1}{\tau} \int_{-\infty}^t \theta(t') e^{(t'-t)/\tau} dt' \right],
  \label{eq:model1}
\end{align}
\end{linenomath*}
where $M_z$ [Eq.~\eqref{eq:Mz}], $r_o$ [Eq.~\eqref{eq:r0}], and $\omega_o$ [Eq.~\eqref{eq:omega0}] are defined in the previous subsection.
The ocean torque is written as the friction coefficient times the difference between the shell rotation rate and the effective ocean angular velocity. A viscoelastic force due to the ice shell drift is denoted $F_e$ and involves a characteristic viscous time scale $\tau$ over which the ice loses its elasticity; this form is similar to the integral form of the Maxwell model \cite{Morrison-2001:understanding}. When the ice viscous adjustment time $\tau$ is very long, the ice remains elastic and $F_e=k_i \theta$, where $k_i$ describes the elastic response of the ice shell and is independent of shell thickness [Eq.~\eqref{eq:ki}]. When $\tau$ is very small, the ice loses its elasticity very quickly, such that in practice, there is no elastic force and $F_e=0$.

Eq.~\eqref{eq:model1} can be converted into the second-order ordinary differential equation \eqref{eq:model2} for the ice drift rate $\omega=d\theta/dt$, which is more convenient to analyze and has the mathematical form of that for a forced and damped harmonic oscillator (except that it is written for the angular velocity rather than for the angle, see Sec.~\ref{sec:sol}). The model's parameters and forcing can be estimated using the current knowledge of Europa's ice shell and results from the ocean simulations (Sections~\ref{sec:ogcm}, \ref{sec:parameters}).

\subsection{Solution for the ice shell drift rate}
\label{sec:sol}
It is easier to solve and analyze the integrodifferential model equation \eqref{eq:model1} after converting it to an ordinary differential equation. This is achieved by taking the time derivative of the second equation of Eq.~\eqref{eq:model1}, using integration by parts and the Leibniz rule. This leads to a single 2nd-order ordinary differential equation where for the angular velocity of the ice shell ($\omega=d\theta/dt$),
\begin{linenomath*}
\begin{align}
    \frac{d^2 \omega}{dt^2}+\left(r_o+\frac1{\tau}\right)  \frac{d\omega}{dt}
    +\left(k_i+\frac{r_o}{\tau}\right)\omega
    =r_o\left(\frac{\omega_o}{\tau}+\frac{d\omega_o}{dt}\right). 
    \label{eq:model2}
\end{align}
\end{linenomath*}
The general solution of the above equation is a sum of a particular solution and the general solution of the homogeneous part. The solution of the homogeneous part of the equation is that of the damped harmonic oscillator (although, again, this equation is for the angular velocity rather than angle as is the case in the standard damped harmonic oscillator), for which $\omega$ decays exponentially in time from its initial conditions. The particular solution of the equation for time-independent forcing, $d\omega_o/dt=0$ is,
\begin{linenomath*}
\begin{align}
    \omega=\frac{r_o\;\omega_o}{k_i\tau+r_o},
  \label{eq:sol1}
\end{align}
\end{linenomath*}
where this is also the steady state solution of the model [Eq.~(\ref{eq:model2})], i.e., $d\omega/dt=0$. Thus, the rate of the ice drift is reduced by the elastic and viscosity parameters $k_i\tau$. Given the decay of the homogeneous solution, the system's initial conditions do not matter beyond the decay time of the homogeneous equation. We note that the time derivative of the forcing $d\omega_o/dt$ in the RHS of Eq.~\eqref{eq:model2} has a very minor role in the solution. 

The frequency spectrum of the ice shell movement for the case of general forcing, $\omega_o(t)$, ignoring the transient decay, can be derived by taking the Fourier transform of Eq.~\eqref{eq:model2} and multiplying by its complex conjugate,
\begin{linenomath*}
\begin{align}
  |\hat{\omega}(\nu)|^2=\frac{|\hat{f}(\nu)|^2}{(k_i+{r_o}/{\tau})^2+(r_o^2+{\tau^{-2}} -2k_i ) \nu^2+\nu^4},
  \label{eq:ps}
\end{align}
\end{linenomath*}
where $|\hat{f}(\nu)|^2$ is the power spectrum of the forcing $r_o\left(\omega_o/{\tau}+{d\omega_o}/{dt}\right)$ and $\nu$ is the frequency. When the forcing is a Gaussian white noise $f(t)=A\xi_t$, $|\hat{f}(\nu)|^2=(2\pi A)^2(1/\tau^2+\nu^2)$ and the power spectrum can be found using Eq.~(\ref{eq:ps}). In this case, the extrema points of the power spectrum can be found where one extremum point is at the zero frequency and the other at,
\begin{linenomath*}
\begin{align}
(\nu^*)^2=-\frac{1}{\tau^2} +k_i \sqrt{1+\frac{2r_o}{k_i \tau}+\frac{2}{k_i \tau^2 }}.
\end{align}
\end{linenomath*}
When the argument of the square root is positive, the maximum of the power spectrum (resonance) is at non-zero frequency; otherwise, the maximum of the power spectrum is at zero, and there is no resonance.

\subsection{Estimating the elastic and viscous parameters of the ice shell}
\label{sec:parameters}
It is possible to estimate the elastic constant of the ice, $k_i$ appearing in our ice shell model \eqref{eq:model1}, based on Eq.~(9) of \citeA{Goldreich-Mitchell-2010:elastic},
\begin{linenomath*}
\begin{align}
    k_i=\frac{24(1+\nu) (1+k_f )^2 q^2 \mu}{5(5+\nu)\rho_i R^2}. \label{eq:ki}
\end{align}
\end{linenomath*}
The definitions and details on the parameters (explanation and estimated value) can be found in  \citeA{Goldreich-Mitchell-2010:elastic} and the estimated value is  $k_i\approx 2.5 \times 10^{-12}$~s$^{-2}$.  Note that in the thin shell limit, $k_i$ does not depend on the shell thickness, because both the elastic torque and the mass of the shell depend linearly on this quantity. If the ice shell is convecting, $k_i$ will be reduced by a factor of the elastic thickness divided by the total shell thickness. This will tend to increase the non-synchronous rotation rate [Eq.~\eqref{eq:sol1}]. The uncertainties on other parameters in this expression are small, except for the rigidity $\mu$. Here we follow  \cite{Goldreich-Mitchell-2010:elastic} and assume that the relevant rigidity is that of intact ice. 

The viscous relaxation (Maxwell) time scale $\tau$ can be estimated by dividing typical ice dynamical viscosity $\eta$ by $\mu$, the ice shell rigidity. The estimated uncertainty range for $\tau$ is quite large. In the case of convecting ice $\eta$ can be estimated as the melting viscosity, which ranges between $10^{13}$--$10^{15}$~Pa~s \cite{Goldsby-Kohlstedt-2001:superplastic}. When the ice is not convecting, the viscosity varies with depth and is much larger near the surface of the ice, due to the lower temperature there.

Calculation of the viscous relaxation timescale, in this case, is not straightforward. In this pilot study, we simply choose an effective upper limit shell viscosity of $10^{17}$~Pa~s to represent the logarithmic mean of the viscosity (near-surface ice is sufficiently cold and brittle that it will not contribute to viscoelastic processes). Less simplistic calculations of the relaxation timescale for realistic ice shell structures should certainly be attempted in the future. For now, we take the overall range of $\eta$ to be $10^{13}-10^{17}$~Pa~s and accordingly
\begin{linenomath*}
\begin{align*}
    \tau =\frac{\eta}{\mu}\sim 2.5\times 10^{3} \rm{s} - 2.5\times 10^{7} \rm{s}.
\end{align*}
\end{linenomath*}

\section{Results}

We consider several scenarios of oceanic forcing on the ice shell: (a) a constant torque, (b) a periodic torque due to the equatorward propagation of Taylor columns outside the tangent cylinder \cite{Ashkenazy-Tziperman-2021:dynamic} that lead to periodic changes in surface ocean currents, (c) a stochastic torque due to the transient nature of the oceanic flow, and (d) an ocean-model-derived forcing. The actual ocean torque (d) is a combination of the first three cases. 

\subsection{Constant ocean forcing}
\begin{figure}[!tbhp]
    \centering
    \includegraphics[width=0.9\textwidth]{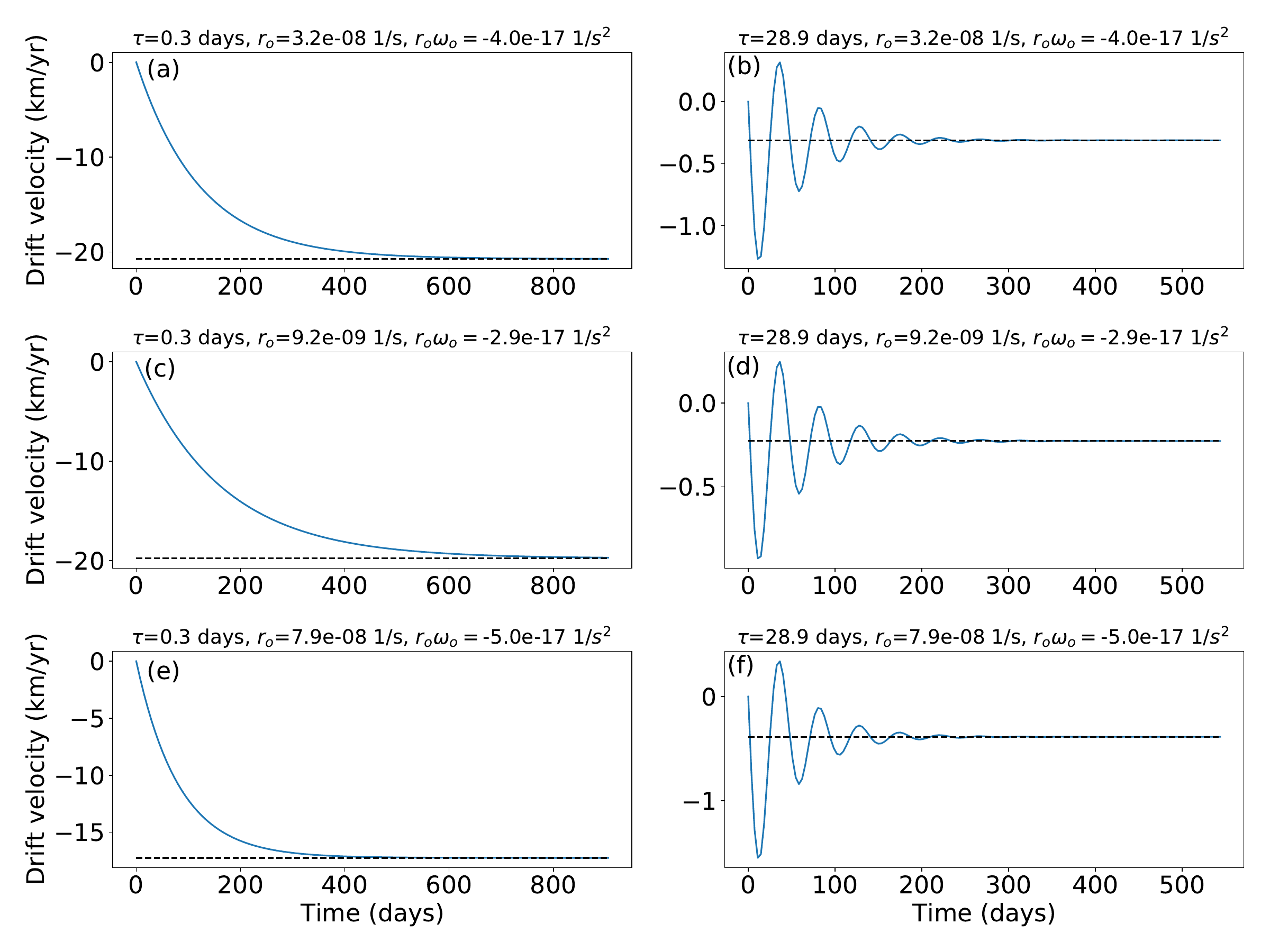}
    \caption{{Time series of the ice drift velocity versus time} for the case of a constant ocean torque, showing the adjustment starting from zero initial conditions. (a) Exponential decay parameter regime for short ice relaxation time scale of 0.3 days, (b) oscillatory regime for long ice relaxation time scale of $\approx$30 days. Panels a and b depict the time series when the linear drag is included in the ocean model (the corresponding ice friction coefficient is $r_o=3.2\times 10^{-8}$~s$^{-1}$), panels c and d depict the time series for the case of ocean simulation without the linear drag (for which the ice friction coefficient is $r_o=9.2\times 10^{-9}$~s$^{-1}$), and panels e and f depict the time series for the case of ocean simulation with large linear drag (for which the ice friction coefficient is $r_o=7.9\times 10^{-8}$~s$^{-1}$). The observational upper bound on the drift velocity is 1 km~yr$^{-1}$ \cite{Hoppa-Greenberg-Geissler-et-al-1999:rotation}.}
    \label{fig:ice-drift-vel}
\end{figure}

Assume first that the effective ocean angular velocity is constant in time (ocean currents are in a steady state). The steady-state solution for the angular velocity of the ice-shell drift, $\omega$, under constant ocean torque forcing (i.e., $\omega_o=\mathrm{const}$) is given in Eq.~\eqref{eq:sol1}.
The ice shell rate of drift is proportional to the ocean forcing $\omega_o$. An increase in $k_i$ or $\tau$ leads to a smaller drift rate; in particular, a longer relaxation timescale (larger $\tau$) would reduce the drift rate. Note that $k_i$ is proportional to ice rigidity $\mu$ while $\eta$ is proportional to $1/\mu$ such that $k_i\tau$ appearing in the denominator of Eq.~\eqref{eq:sol1} should be independent of $\mu$. When starting from arbitrary initial conditions, the adjustment to the constant drift solution (Fig.~\ref{fig:ice-drift-vel}) involves either an exponential decay to the steady state or oscillations whose amplitude decays exponentially.  Oscillations occur (Fig.~\ref{fig:ice-drift-vel}b,d,f) when the time scale $\tau$ is large, with a frequency $\sqrt{k_i-(r_o+1/\tau)^2/4}$; when the argument under the square root is negative (for small $\tau$), exponential decay occurs (Fig.~\ref{fig:ice-drift-vel}a,c,e). This damped internal oscillatory behavior plays an important role when the ocean forcing is periodic or stochastic, as discussed next.

\begin{figure}[!tbhp]
    \centering
    \includegraphics[width=0.9\textwidth]{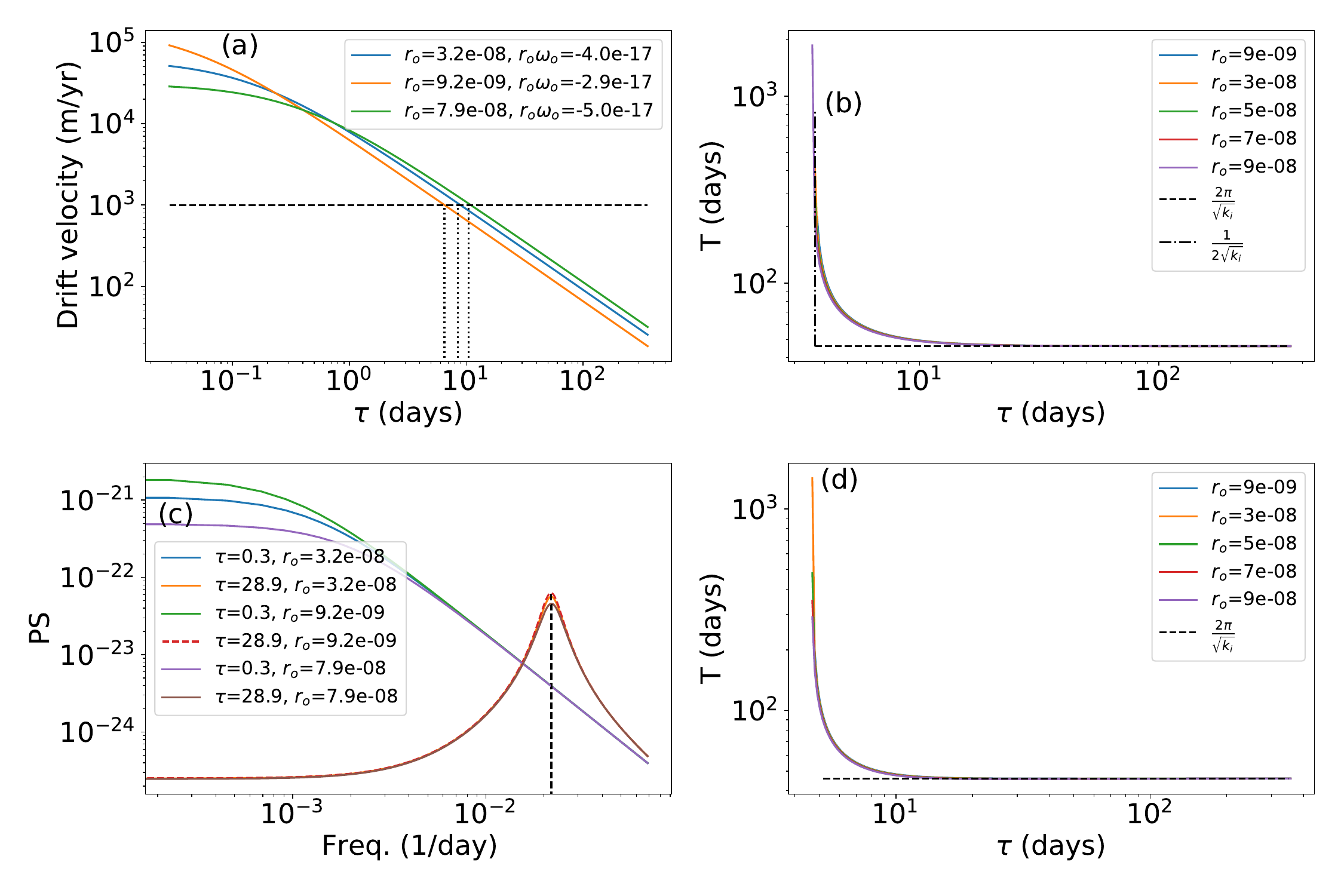}
    \caption{Parameter sensitivity. (a) Steady-state solution for drift velocity as a function of $\tau$ for a constant ocean torque for three cases, with (blue) standard oceanic drag ($\tilde{r}_o=2\times10^{-4}$ m~s$^{-1}$, $r=3.2\times10^{-8}$ 1~s$^{-1}$), without (orange) oceanic linear drag ($\tilde{r}_o=0$, $r=9.2\times10^{-9}$ 1~s$^{-1}$), and with large (green) ocean linear drag ($\tilde{r}_o=6\times10^{-4}$ m~s$^{-1}$, $r=7.9\times10^{-8}$ s$^{-1}$). The corresponding oceanic torque $r_o\omega_o$ is indicated in the figure legend. The dashed line shows the observationally constrained upper limit of the ice shell drift velocity of 1 km~yr$^{-1}$ \cite{Hoppa-Greenberg-Geissler-et-al-1999:rotation}, indicating that $\tau$ should be larger than $\sim$$10$ days, for all cases; the vertical dotted lines indicate the minimal $\tau$ of the different cases. (b) Oscillation period ($T$, see Fig.~\ref{fig:ice-drift-vel}b) as a function of $\tau$ for a constant ocean torque and five friction coefficient, $r_o$, values. The oscillation period converges to $\sim$$45$ days for realistic $\tau$ that is larger than $\sim$$10$ days. (c) Power spectra of the drift velocity as a function of frequency for different model parameters for the stochastic forcing case, corresponding to the cases shown in panel a. (d) Period corresponding to the spectral peak under stochastic forcing as a function of $\tau$. Also here, the period converges to $\sim$$45$ days, as for the constant forcing case shown in panel b.}
    \label{fig:ice-drift-ve-and-noise}
\end{figure}

The shell relaxation time scale, $\tau$, can drastically affect the ice-drift velocity, as its uncertainty spans at least four orders of magnitude (from one hour to hundreds of days). The resulting ice shell drift velocity ranges from a few tens of meters per year to almost one hundred km per year (Fig.~\ref{fig:ice-drift-ve-and-noise}a). Observational constraints \cite{Hoppa-Greenberg-Geissler-et-al-1999:rotation} indicate that the ice-shell velocity is smaller than 1 km~yr$^{-1}$ (dashed line in Fig~\ref{fig:ice-drift-ve-and-noise}a) and this upper limit constrains the shell relaxation time scale to be larger than about ten days, for all simulations of different oceanic linear drag. Conversely, unless the effective ice shell viscosity exceeds $10^{17}$~Pa~s, which would only occur for a non-convecting shell \cite{McKinnon-1999:convective}, we expect a non-synchronous rotation period of $\lesssim 3\times10^5$~yrs (Fig.~\ref{fig:ice-drift-ve-and-noise}a) or a drift velocity larger than about 30~m~yr$^{-1}$. Such a drift period would be readily detectable with a future mission to Europa.

In Fig.~\ref{fig:ice-drift-ve-and-noise}b, we present the period of the internal oscillations of the ice shell (i.e., $2\pi/\sqrt{k_i-(r_o+1/\tau)^2/4}$) as a function of the shell relaxation time scale, $\tau$, and ocean friction parameter, $r_o$. We show that $r_o$ hardly affects the period for realistic values of $r_o$ while the relaxation time scale, $\tau$, can have a much larger effect on the period. Given the constraint that $\tau\gtrsim 10$ days, the oscillation period (Fig.~\ref{fig:ice-drift-ve-and-noise}b) is nearly independent of the actual value of $\tau$ and converges to $T\sim2\pi/\sqrt{k_i}\approx 45$ days; {the oscillation period depends mainly on the shell rigidity, while the mean drift rate depends on $\tau$.}

\subsection{Periodic forcing} We assume, for simplicity, that $\omega_o(t)$ is a pure cosine function. The solution of Eq.~\eqref{eq:model2} is periodic with the same period as the forcing, and resonance is obtained when the forcing frequency equals $(\nu^*)^2=-1/\tau^2 +k_i \sqrt{1+(2r_o)/(k_i \tau)+2/(k_i \tau^2 )}$. For large enough $\tau$, $\nu^*=\sqrt{k_i}$. Note that the resonance frequency is related to, but not identically equal to, the frequency of the damped internal oscillations given above. 

\subsection{Stochastic forcing} The power spectrum of the drift can be evaluated for a general forcing [Eq.~\eqref{eq:ps}]. There are two types of responses to a white noise forcing, one with a spectral peak at the resonance frequency $\nu^*$ (on the right side of Fig.~\ref{fig:ice-drift-ve-and-noise}c), and another with a monotonically decreasing Lorentzian-like power spectrum with increasing frequency, showing a transition from a plateau at low frequencies to a power-law decay for higher frequencies. The transition (crossover) point ($\sim1000$ days) indicates the expected time scale of the forced stochastic variability of the ice shell drift. As before, the numerical value of the ocean friction coefficient, $r_o$, does not significantly affect the spectra within its range of uncertainty, for realistic values of $\tau$. Variations of the ice relaxation time scale $\tau$ within its own range of uncertainty results in the above two different types of spectra. In Fig.~\ref{fig:ice-drift-ve-and-noise}d we plot the period at the resonance frequency for the uncertainty range of $\tau$ and $r_o$. Different values of $\tau$ can result in very different periods of the spectral peak, from about 50 days to 1000 days. Since $\tau$ should be larger than about ten days to satisfy the ice drift constraint of \citeA{Hoppa-Greenberg-Geissler-et-al-1999:rotation}, one expects the spectral peak of the ice shell drift rate oscillations due to white noise stochastic forcing to be around 45 days. The expected magnitude of these oscillations depends on the magnitude of the stochastic forcing; as discussed below in the context of the solution driven by the ocean model solution, the oscillations are expected to be small.

\subsection{The ocean-model-derived forcing case} 
\begin{figure}[!tbhp]
    \centering
    \includegraphics[width=0.9\textwidth]{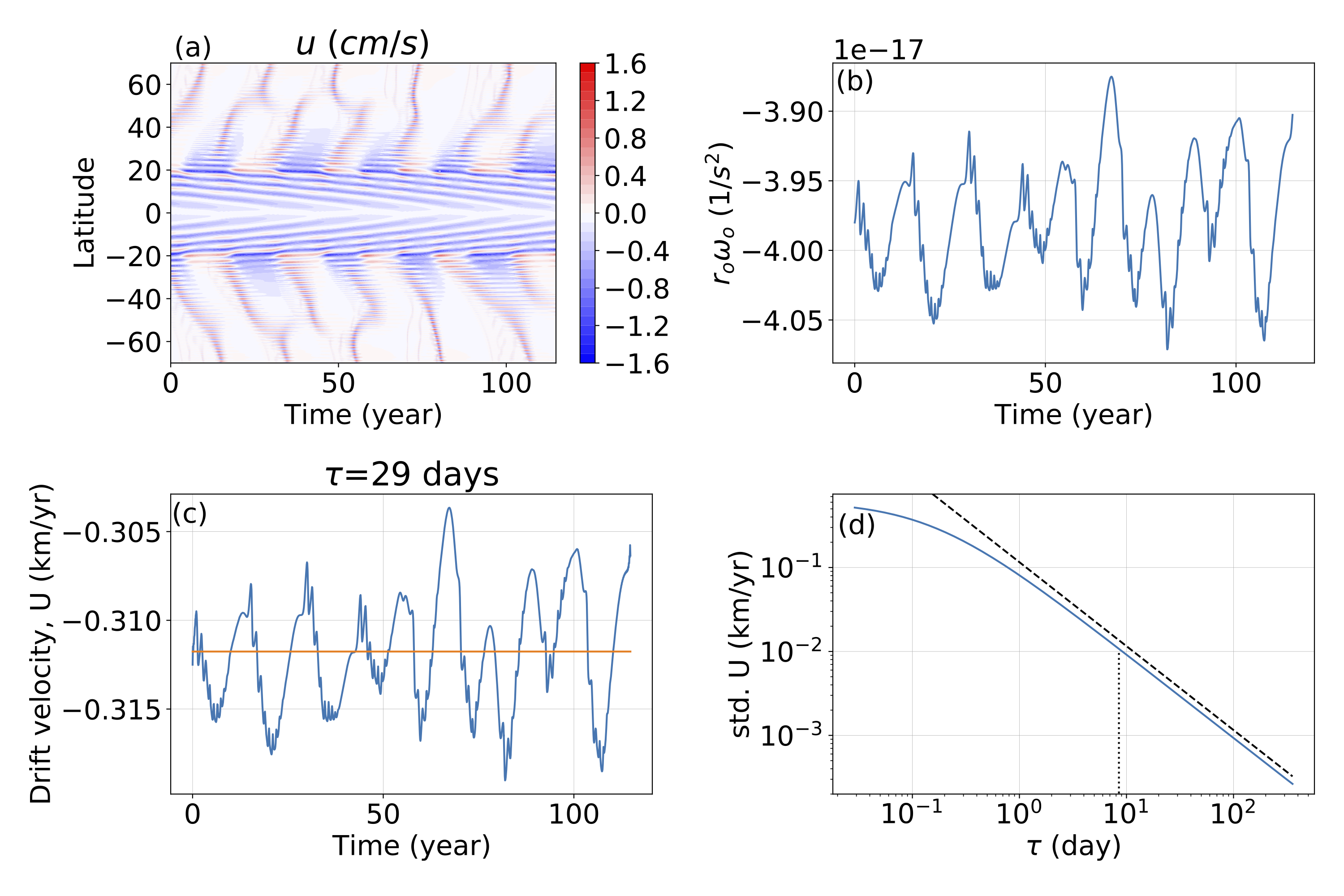}
    \caption{{Ice shell drift forcing and response:} (a) The ocean zonal velocity (cm/s) as a function of time (in years) and latitude. Note the difference between the equatorward propagating Taylor columns outside the tangent cylinder (20\degree{S}--20\degree{N}) and the periodic pattern within the tangent cylinder \cite{Ashkenazy-Tziperman-2021:dynamic}. Note also the westward flow of the equatorial current. (b) The global average of the zonal torque due to ocean currents, $r_o\omega_o$, as a function of time.
      (c) Blue line: drift velocity of a point on the equator of Europa (km/year). Orange line: the ice-shell drift velocity when forcing the model by the constant temporal mean ocean torque. Here $\tau$=29 days. (d) The standard deviation of the ice-shell drift velocity as a function of the relaxation time scale $\tau$. The std decreases like $1/\tau$ (dashed line) for large $\tau$.}
    \label{fig:u-and-zonally-averaged-u}
\end{figure}

Fig.~\ref{fig:u-and-zonally-averaged-u}c shows the ice-shell drift velocity for the ocean torque forcing derived from our ocean model, shown in Fig.~\ref{fig:u-and-zonally-averaged-u}b; linear oceanic drag was included in this simulation (corresponding to an ice model friction coefficient of $r_o=3.2\times 10^{-8}$~s$^{-1}$). The results for the oceanic simulation in the absence of linear oceanic drag (corresponding to an ice model friction coefficient of $r_o=9.2\times 10^{-9}$~s$^{-1}$) and with a large linear oceanic drag (corresponding to an ice model friction coefficient of $r_o=7.9\times 10^{-8}$~s$^{-1}$) are presented in
Supplementary Figs.~S1, S2. This solution combines features from the above-mentioned constant, periodic, and stochastic forcing scenarios. The power spectrum of the forcing (Supplementary
Fig.~S3) shows a strong peak at about ten years resulting from the speed of propagation of the Taylor columns, indicating that the forcing is nearly periodic. This $\sim$10~yr period is much larger than the resonance period of $\sim45$ days discussed above (Fig.~\ref{fig:ice-drift-ve-and-noise}d), and we, therefore, do not expect a resonant response or a peak in the NSR drift spectrum due to the ocean current forcing. Moreover, the noisy fluctuations superimposed on the nearly-periodic ocean torque forcing are relatively small (as seen by the tails of the 10-year peak of the power spectrum). As a result, a resonance due to this noisy part of the forcing is likely to be negligible relative to the response to the oscillatory part of the forcing. As $\tau$ increases, the standard deviation of the ice shell drift fluctuations driven by the deviations of the ocean model torque from its long-term mean decreases like $1/\tau$ (dashed line in Fig.~\ref{fig:u-and-zonally-averaged-u}d). Given the observational constraint of $\tau>10$ days, Fig.~\ref{fig:u-and-zonally-averaged-u}d shows that the ice shell movements due to this time-variable ocean forcing are expected to be very small (standard deviation $<10$ m~yr$^{-1}$). Since the drift rate, $\omega$ [Eq.~\eqref{eq:sol1}] is proportional to $1/\tau$ (i.e., $\omega\approx{}r_o\omega_o/(k_i\tau)$ for $k_i\tau\gg r_o$), the ratio between the standard deviation and the mean of the drift rate $\omega$ is expected to remain constant for large $\tau$. We conclude that the time-averaged ocean torque is expected to dominate the ice-shell drift rate, resulting in steady drift. Since $\tau \gtrsim$10~days ($\sim 10^6$~s) is required by the observations (Fig.~\ref{fig:ice-drift-ve-and-noise}a), the effective ice shell viscosity must exceed about $\rm 3\times10^{15}$~Pa~s. Although this effective viscosity will only roughly correspond to the interior viscosity of a convecting ice shell, a value of $\rm 3\times10^{15}~Pa~s$ may imply that at best sluggish convection is occurring \cite{McKinnon-1999:convective}.

\section{Discussions}
A surprising result of our model is that the torque calculated by the ocean model is negative and thus decreases the ice shell rotation rate relative to the synchronous rotation case. This is in contrast to the effect of tidal forcing which tends to increase the rotation rate  \cite{Hoppa-Greenberg-Geissler-et-al-1999:rotation}. It has been shown that tectonic crack orientations can equally well be explained by retrograde NSR as by prograde NSR  \cite{Sarid-Greenberg-Hoppa-et-al-2004:crack}, so either possibility is currently viable based on existing observations. 

One way of detecting NSR is to look for offsets between the predicted and observed location of the terminator (the line separating the daylight and night side) \cite{Hoppa-Greenberg-Geissler-et-al-1999:rotation}. Some of the best {\em Galileo} images of the terminator have a resolution of 0.4~km \cite{Hoppa-Greenberg-Geissler-et-al-1999:rotation}. Future spacecraft observations are likely to have resolutions significantly better than this. Assuming a time interval of 30~years between the {\em Galileo} observations and future imaging campaigns, a terminator location with a precision of 0.4~km implies that an NSR period of about $7\times10^5$~years (drift rate of 15~m~yr$^{-1}$) would be marginally detectable, larger than our estimate for the slowest ice-shell drift of $\sim$30~m~yr$^{-1}$.

The model we developed for the ice shell drift is highly simplified. Moreover, it relies on the oceanic general circulation model we use, whose parameters are partially based on Earth's ocean parameters (like the vertical viscosity and linear drag coefficient) and hence uncertain in the context of Europa. While we have demonstrated that the general conclusion of retrograde drift is insensitive to some of the main model parameters (like the linear drag), one cannot rule out the possibility that sensitivity to the ocean model setup and other parameters (such as the distribution of bottom heating) won't cause an opposite drift, implying a non-negligible uncertainty in the magnitude and sign of our estimated drift velocity.

Another effect that was not included in the proposed model is the effect of a tidal phase lag on the drift of the ice shell. Such a lag would impose an additional torque tending to spin the satellite up \cite{Goldreich-Mitchell-2010:elastic}. However, the size of the lag for Europa is expected to be small, $\rm < 0.3^\circ$ for a shell thickness $<$30~km \cite{Moore-Schubert-2000:tidal}, and as a result, we estimate this torque to be smaller than the ocean torque, although certainly not negligible. See the Supplementary Material for an estimate of this torque and additional discussion.

Further work is needed to understand whether tidal or ocean torques will dominate. 3D simulations of Europa {without linear drag} are highly turbulent \cite{Ashkenazy-Tziperman-2021:dynamic} {and result in super-rotation (eastward flow) at the upper, low latitude, ocean; yet, our preliminary 3D simulation with linear drag indicates that the flow is negative (westward)}. Moreover, this preliminary simulation exhibits a periodicity of $\sim$50 days (most probably due to eddy and convection dynamics) which may resonate with the internal ice dynamics reported here. The ocean eddies are expected to add to the amplitude and change the spectral characteristics of the stochastic forcing experienced by the ice shell. The computational cost of 3D eddy-resolving simulations prohibits obtaining a sufficiently long time series of ocean torques to be used to drive the ice shell model drift equation used here. Apart from our use of a 2D ocean model, other significant assumptions include a single viscoelastic adjustment time scale and neglect of tidal forcing. Our drift rate predictions, which depend primarily on the viscoelastic relaxation timescale $\tau$ (Fig.~\ref{fig:ice-drift-ve-and-noise}a) are necessarily uncertain because of uncertainty in $\tau$. Yet the constraint of $\tau \gtrsim 10$~days indicates that the effective ice shell viscosity should be $\eta \gtrsim 3 \times 10^{15}$ Pa s. 

\section{Conclusions} The model suggested here demonstrates how the combination of ocean torques due to zonal surface currents and viscous adjustment of the ice shell may lead to a potentially observable, retrograde non-synchronous drift (Fig.~\ref{fig:ice-drift-ve-and-noise}a). Oscillatory or stochastic rotation rate variations are expected to be minor, ruling out potential confusion with long-period librations \cite{Rambaux-Hoolst-Karatekin-2011:librational}. Future missions and observations are expected to lead to an improved constraint on the ice shell drift velocity. The expected precision is about 15~m~yr$^{-1}$ or better, smaller than our anticipated slowest drift rate of 30~m~yr$^{-1}$. Using the model developed here, such closer bounds can lead to a tighter constraint on the viscous response time of the ice shell, which plays a dominant role here. Conversely, a tighter constraint on this ice shell viscous response due to a better understanding of the physics affecting this time scale (ice crystal size, brittle failure, the temperature distribution within the ice, etc.) would permit inferences of Europa's ocean currents.



\acknowledgments
We thank Roiy Sayag for helpful discussions. This work was funded by the Bi-National US-Israel Science Foundation (grant no. 2018152), ET was funded by DOE grant DE-SC0023134, and thanks the Weizmann Institute for its hospitality during parts of this work. 

\section*{Conflict of Interest}
The authors declare no conflicts of interest relevant to this study.

\section*{Data Availability Statement}

The model presented in this paper is based on surface ocean currents that were simulated using the MITgcm \cite{MITgcm-manual-github:mitgcm}. Additional information regarding these simulations can be found in \citeA{Ashkenazy-Tziperman-2021:dynamic} and the setup files of the numerical simulations can be downloaded from \texttt{https://doi.org/10.17605/OSF.IO/SVXBQ}. The simulated upper ocean zonal currents presented in Fig.~\ref{fig:u-and-zonally-averaged-u}a can be downloaded from \texttt{https://osf.io/svxbq/files/osfstorage/63d655ad34869301e90b5505}.

\newpage

\section*{Supplementary Material}

\subsection*{Estimating the tidal torque}

    For a satellite in a non-circular orbit, there will be a net tidal torque acting to spin the satellite up \cite{Greenberg-Weidenschilling-1984:how}. The magnitude of this torque depends on the lag angle between Europa's tidal bulge and Jupiter, which in turn depends on the viscoelastic properties of the satellite. The model derivation in Section~\ref{sec:icy-shell-model} assumes the lag angle to be small and therefore neglects the tidal torque due to Europa's eccentricity. In the case of a decoupled shell, the tidal torque averaged over the orbit based on equation 16 from \citeA{Goldreich-Mitchell-2010:elastic} is given by
    \begin{align*}
        T_{e-tide}=\frac{576}{5}\pi\left(\frac{1+\nu}{5+\nu}\right)
        \frac{h_2^2}{Q}(qe)^2\mu{}dR^2.
    \end{align*}
    Here, $\nu\approx 0.3$ is the Poisson's ratio, $h_2=1.25$ is the Love number representing the amplitude of tidal deformation, $Q\approx1/2\delta$ is a measure of the ice shell phase lag, $\delta$ is the angle between the bulge and Jupiter, $q=5.4\times10^{-4}$ the ratio of the centripetal to gravitational acceleration, $e=0.01$ the orbital eccentricity, $\mu$ the rigidity, $d$ the ice shell thickness and $R$ Europa's radius. We take $\mu$=3~GPa for ice, $d$=10~km and $\delta < 0.3^\circ$ based on the results of \citeA{Moore-Schubert-2000:tidal} for $d<$30~km.  The resulting estimate of the net torque is $<3\times10^{15}$~N~m. This may be compared with typical ocean torques, which from Fig~\ref{fig:u-and-zonally-averaged-u}b are roughly $2\times10^{16}$~N~m. So the tidal torques are not negligible but would only contribute at the $< 15\%$ level. The ice shell model formulation is more complicated when attempting to take into account a finite value of the bulge angle, so we elected to neglect this contribution as a first attempt at estimating the shell drift due to the ocean torque.

\setcounter{figure}{0}
\renewcommand{\thefigure}{S\arabic{figure}}

\begin{figure}[!tbhp]
    \centering
    \includegraphics[width=0.9\textwidth]{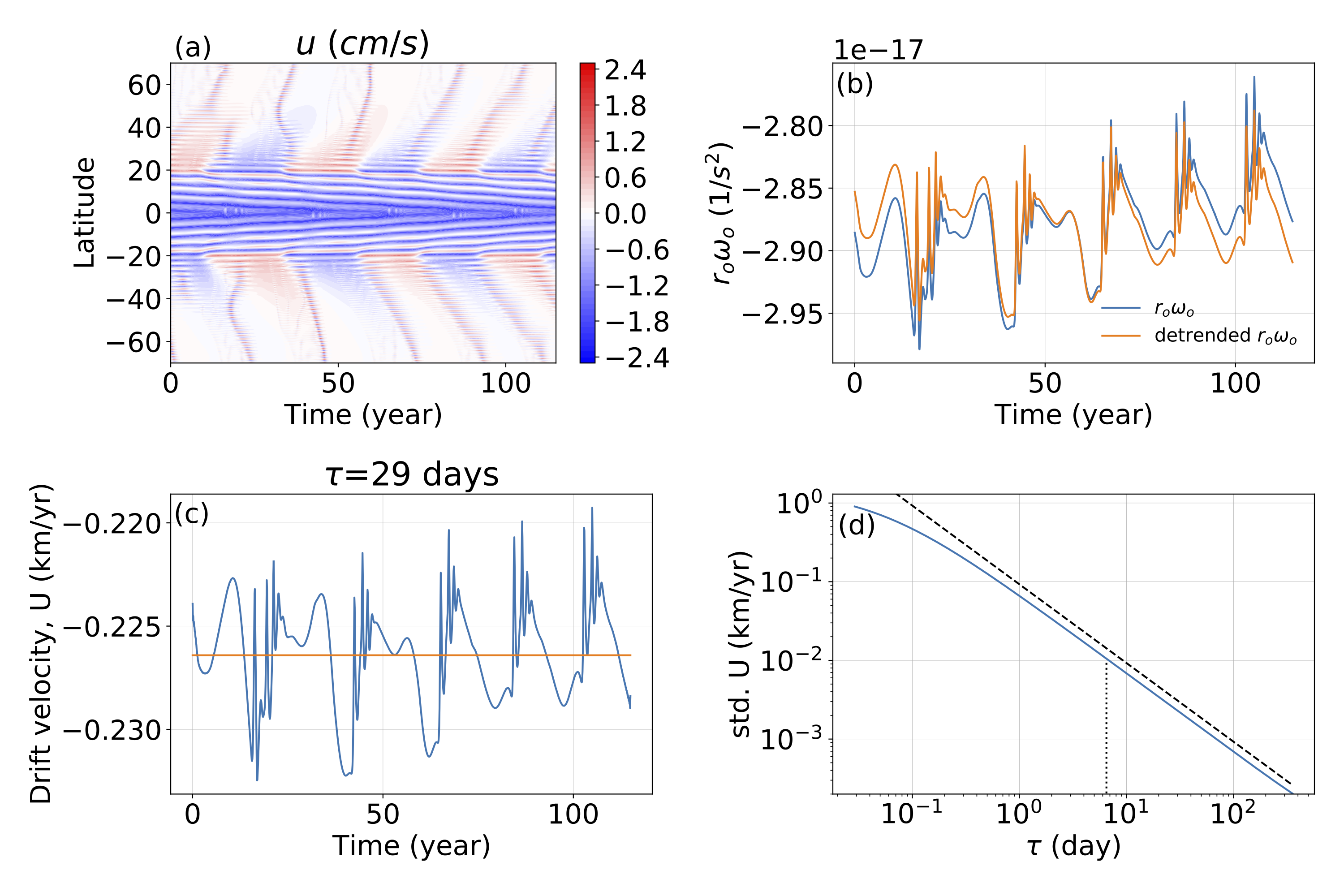}
    \caption{{ice shell drift forcing and response when linear drag is not included in the ocean model.} Same as Fig.~\ref{fig:u-and-zonally-averaged-u} for the ocean simulation in which linear drag is not included. In this case $r_o=9.2\times 10^{-9}$ s$^{-1}$ and $r_o\omega_o= -2.9\times 10^{-17}$ s$^{-2}$. In panel a, the original time series exhibits oscillations that are superimposed on a weak trend (blue). The detrended time series is indicated by the orange line and is used to find the ice-shell drift velocity shown in panel (c).}
    \label{fig:u-and-zonally-averaged-u-no-linear-drag}
\end{figure}

\begin{figure}[!tbhp]
    \centering
    \includegraphics[width=0.9\textwidth]{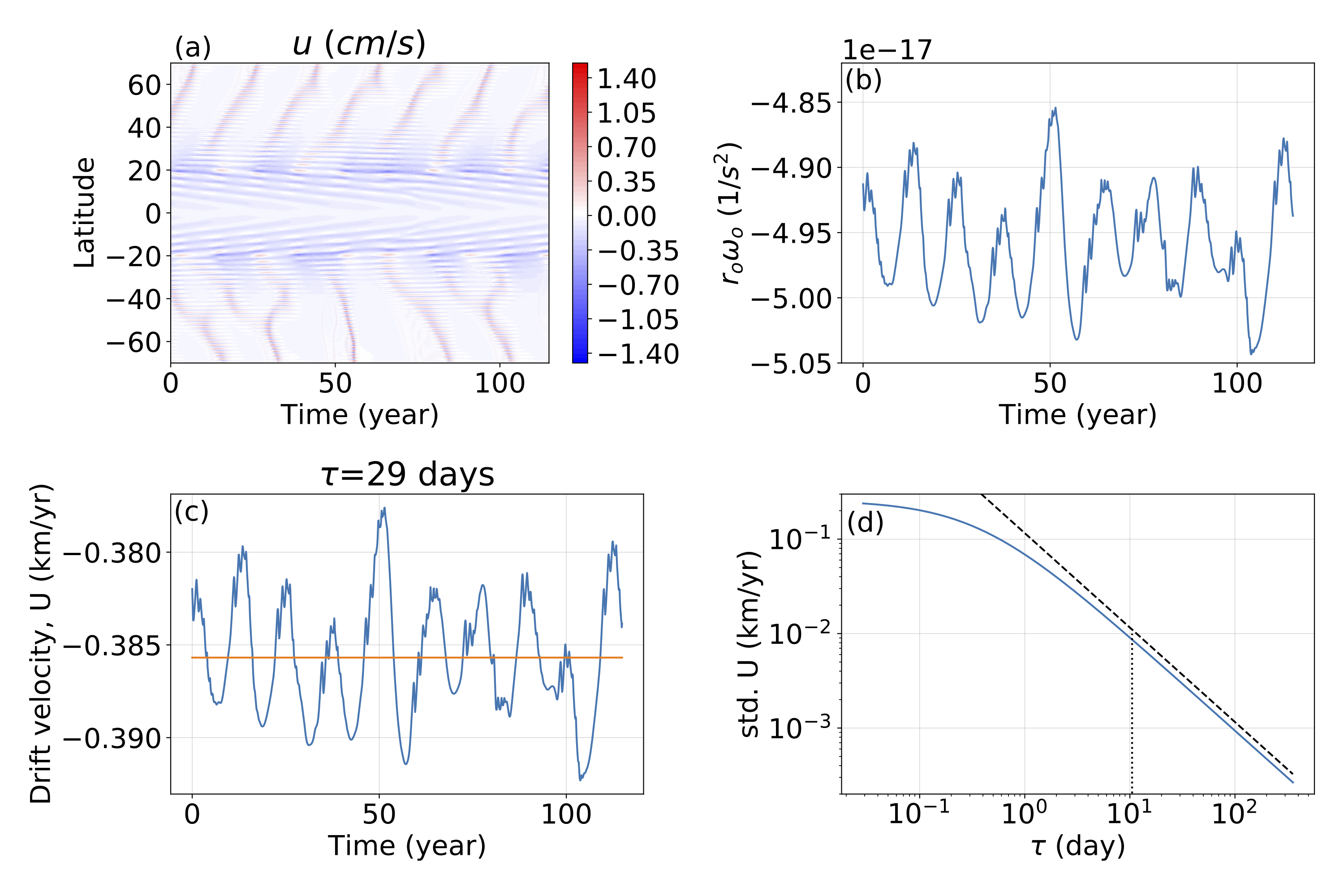}
    \caption{{ice shell drift forcing and response when high linear drag ($6\times10^{-4}$m~s$^{-1}$) is included in the ocean model.} Same as Fig.~\ref{fig:u-and-zonally-averaged-u} for the ocean simulation when using linear drag coefficient that is three times larger than a typical drag coefficient. In this case $r_o=7.9\times 10^{-8}$ s$^{-1}$ and $r_o\omega_o=-5.0\times 10^{-17}$ s$^{-2}$. 
}
    \label{fig:u-and-zonally-averaged-u-high-linear-drag}
\end{figure}

\begin{figure}[!tbhp]
    \centering
    \includegraphics[width=0.9\textwidth]{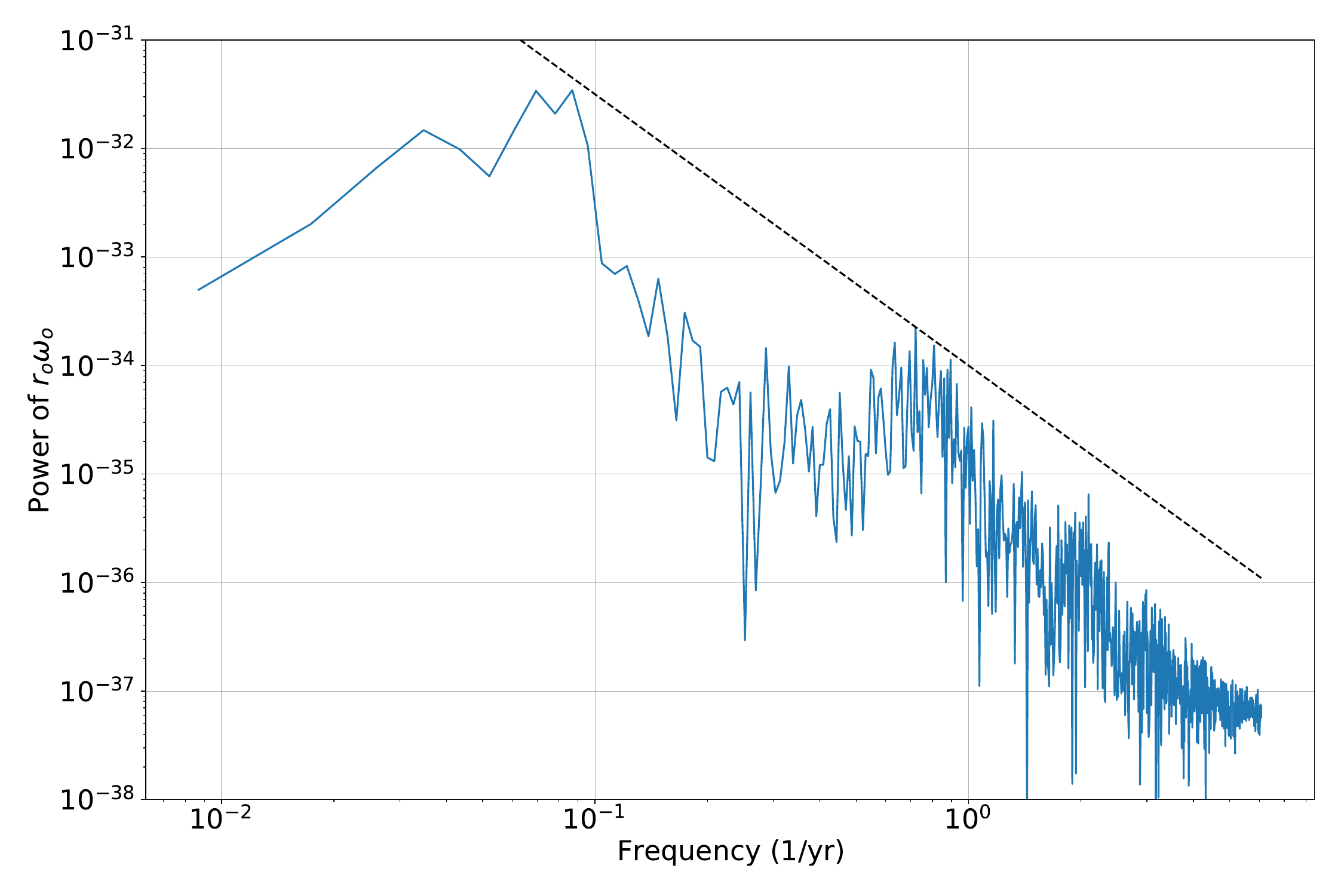}
    \caption{The power spectrum of the forcing torque $r_o\omega_o$ shown in Fig.~\ref{fig:u-and-zonally-averaged-u}b shows the dominant frequency of $\sim$10 years superimposed on a power-law (dashed line) noise tail ($\propto \nu^{-2.5}$ where $\nu$ is the frequency). }
    \label{fig:torque_power_spectrum}
\end{figure}

\newpage


\begin{thebibliography}{}

\bibitem [\protect \citeauthoryear {%
Ashkenazy%
}{%
Ashkenazy%
}{%
{\protect \APACyear {2018}}%
}]{%
Ashkenazy-2019:surface}
\APACinsertmetastar {%
Ashkenazy-2019:surface}%
\begin{APACrefauthors}%
Ashkenazy, Y.%
\end{APACrefauthors}%
\unskip\
\newblock
\APACrefYearMonthDay{2018}{}{}.
\newblock
{\BBOQ}\APACrefatitle {The surface temperature of {Europa}} {The surface
  temperature of {Europa}}.{\BBCQ}
\newblock
\APACjournalVolNumPages{Heliyon}{5}{e01908}{arxiv.org/abs/1608.07372}.
\PrintBackRefs{\CurrentBib}

\bibitem [\protect \citeauthoryear {%
Ashkenazy%
, Sayag%
\BCBL {}\ \BBA {} Tziperman%
}{%
Ashkenazy%
\ \protect \BOthers {.}}{%
{\protect \APACyear {2018}}%
}]{%
Ashkenazy-Sayag-Tziperman-2018:dynamics}
\APACinsertmetastar {%
Ashkenazy-Sayag-Tziperman-2018:dynamics}%
\begin{APACrefauthors}%
Ashkenazy, Y.%
, Sayag, R.%
\BCBL {}\ \BBA {} Tziperman, E.%
\end{APACrefauthors}%
\unskip\
\newblock
\APACrefYearMonthDay{2018}{}{}.
\newblock
{\BBOQ}\APACrefatitle {Dynamics of the global meridional ice flow of {Europa}'s
  icy shell} {Dynamics of the global meridional ice flow of {Europa}'s icy
  shell}.{\BBCQ}
\newblock
\APACjournalVolNumPages{Nature Astronomy}{2}{1}{43}.
\PrintBackRefs{\CurrentBib}

\bibitem [\protect \citeauthoryear {%
Ashkenazy%
\ \BBA {} Tziperman%
}{%
Ashkenazy%
\ \BBA {} Tziperman%
}{%
{\protect \APACyear {2021}}%
}]{%
Ashkenazy-Tziperman-2021:dynamic}
\APACinsertmetastar {%
Ashkenazy-Tziperman-2021:dynamic}%
\begin{APACrefauthors}%
Ashkenazy, Y.%
\BCBT {}\ \BBA {} Tziperman, E.%
\end{APACrefauthors}%
\unskip\
\newblock
\APACrefYearMonthDay{2021}{}{}.
\newblock
{\BBOQ}\APACrefatitle {Dynamic {Europa} ocean shows transient {Taylor} columns
  and convection driven by ice melting and salinity} {Dynamic {Europa} ocean
  shows transient {Taylor} columns and convection driven by ice melting and
  salinity}.{\BBCQ}
\newblock
\APACjournalVolNumPages{Nature Communications}{12}{1}{1--12}.
\PrintBackRefs{\CurrentBib}

\bibitem [\protect \citeauthoryear {%
Billings%
\ \BBA {} Kattenhorn%
}{%
Billings%
\ \BBA {} Kattenhorn%
}{%
{\protect \APACyear {2005}}%
}]{%
Billings-Kattenhorn-2005:great}
\APACinsertmetastar {%
Billings-Kattenhorn-2005:great}%
\begin{APACrefauthors}%
Billings, S\BPBI E.%
\BCBT {}\ \BBA {} Kattenhorn, S\BPBI A.%
\end{APACrefauthors}%
\unskip\
\newblock
\APACrefYearMonthDay{2005}{}{}.
\newblock
{\BBOQ}\APACrefatitle {The great thickness debate: {Ice} shell thickness models
  for {Europa} and comparisons with estimates based on flexure at ridges} {The
  great thickness debate: {Ice} shell thickness models for {Europa} and
  comparisons with estimates based on flexure at ridges}.{\BBCQ}
\newblock
\APACjournalVolNumPages{Icarus}{177}{2}{397--412}.
\PrintBackRefs{\CurrentBib}

\bibitem [\protect \citeauthoryear {%
Carr%
\ \protect \BOthers {.}}{%
Carr%
\ \protect \BOthers {.}}{%
{\protect \APACyear {1998}}%
}]{%
Carr-Belton-Chapman-et-al-1998:evidence}
\APACinsertmetastar {%
Carr-Belton-Chapman-et-al-1998:evidence}%
\begin{APACrefauthors}%
Carr, M\BPBI H.%
, Belton, M\BPBI J.%
, Chapman, C\BPBI R.%
, Davies, M\BPBI E.%
, Geissler, P.%
, Greenberg, R.%
\BDBL {}others%
\end{APACrefauthors}%
\unskip\
\newblock
\APACrefYearMonthDay{1998}{}{}.
\newblock
{\BBOQ}\APACrefatitle {Evidence for a subsurface ocean on {Europa}} {Evidence
  for a subsurface ocean on {Europa}}.{\BBCQ}
\newblock
\APACjournalVolNumPages{Nature}{391}{6665}{363--365}.
\PrintBackRefs{\CurrentBib}

\bibitem [\protect \citeauthoryear {%
Cassen%
, Reynolds%
\BCBL {}\ \BBA {} Peale%
}{%
Cassen%
\ \protect \BOthers {.}}{%
{\protect \APACyear {1979}}%
}]{%
Cassen-Reynolds-Peale-1979:is}
\APACinsertmetastar {%
Cassen-Reynolds-Peale-1979:is}%
\begin{APACrefauthors}%
Cassen, P.%
, Reynolds, R\BPBI T.%
\BCBL {}\ \BBA {} Peale, S.%
\end{APACrefauthors}%
\unskip\
\newblock
\APACrefYearMonthDay{1979}{}{}.
\newblock
{\BBOQ}\APACrefatitle {Is there liquid water on {Europa}?} {Is there liquid
  water on {Europa}?}{\BBCQ}
\newblock
\APACjournalVolNumPages{Geophys. Res. Lett.}{6}{9}{731--734}.
\PrintBackRefs{\CurrentBib}

\bibitem [\protect \citeauthoryear {%
Chapman%
, Merline%
, Bierhaus%
, Brooks%
\BCBL {}\ \BBA {} Team%
}{%
Chapman%
\ \protect \BOthers {.}}{%
{\protect \APACyear {1998}}%
}]{%
Chapman-Merline-Bierhaus-et-al-1998:cratering}
\APACinsertmetastar {%
Chapman-Merline-Bierhaus-et-al-1998:cratering}%
\begin{APACrefauthors}%
Chapman, C.%
, Merline, W.%
, Bierhaus, B.%
, Brooks, S.%
\BCBL {}\ \BBA {} Team, G\BPBI I.%
\end{APACrefauthors}%
\unskip\
\newblock
\APACrefYearMonthDay{1998}{}{}.
\newblock
{\BBOQ}\APACrefatitle {Cratering in the jovian system: {Intersatellite}
  comparisons} {Cratering in the jovian system: {Intersatellite}
  comparisons}.{\BBCQ}
\newblock
\BIn{} \APACrefbtitle {Lunar and Planetary Science Conference} {Lunar and
  planetary science conference}\ (\BPG~1927).
\PrintBackRefs{\CurrentBib}

\bibitem [\protect \citeauthoryear {%
Chyba%
\ \BBA {} Phillips%
}{%
Chyba%
\ \BBA {} Phillips%
}{%
{\protect \APACyear {2001}}%
}]{%
Chyba-Phillips-2001:possible}
\APACinsertmetastar {%
Chyba-Phillips-2001:possible}%
\begin{APACrefauthors}%
Chyba, C\BPBI F.%
\BCBT {}\ \BBA {} Phillips, C\BPBI B.%
\end{APACrefauthors}%
\unskip\
\newblock
\APACrefYearMonthDay{2001}{}{}.
\newblock
{\BBOQ}\APACrefatitle {Possible ecosystems and the search for life on {Europa}}
  {Possible ecosystems and the search for life on {Europa}}.{\BBCQ}
\newblock
\APACjournalVolNumPages{Proc. Natl. Acad. Sci. U.S.A.}{98}{3}{801--804}.
\PrintBackRefs{\CurrentBib}

\bibitem [\protect \citeauthoryear {%
Geissler%
\ \protect \BOthers {.}}{%
Geissler%
\ \protect \BOthers {.}}{%
{\protect \APACyear {1998}}%
}]{%
Geissler-Greenberg-Hoppa-et-al-1998:evidence}
\APACinsertmetastar {%
Geissler-Greenberg-Hoppa-et-al-1998:evidence}%
\begin{APACrefauthors}%
Geissler, P.%
, Greenberg, R.%
, Hoppa, G.%
, Helfenstein, P.%
, McEwen, A.%
, Pappalardo, R.%
\BDBL {}others%
\end{APACrefauthors}%
\unskip\
\newblock
\APACrefYearMonthDay{1998}{}{}.
\newblock
{\BBOQ}\APACrefatitle {Evidence for non-synchronous rotation of {Europa}}
  {Evidence for non-synchronous rotation of {Europa}}.{\BBCQ}
\newblock
\APACjournalVolNumPages{Nature}{391}{6665}{368--370}.
\PrintBackRefs{\CurrentBib}

\bibitem [\protect \citeauthoryear {%
Gissinger%
\ \BBA {} Petitdemange%
}{%
Gissinger%
\ \BBA {} Petitdemange%
}{%
{\protect \APACyear {2019}}%
}]{%
Gissinger-Petitdemange-2019:magnetically}
\APACinsertmetastar {%
Gissinger-Petitdemange-2019:magnetically}%
\begin{APACrefauthors}%
Gissinger, C.%
\BCBT {}\ \BBA {} Petitdemange, L.%
\end{APACrefauthors}%
\unskip\
\newblock
\APACrefYearMonthDay{2019}{}{}.
\newblock
{\BBOQ}\APACrefatitle {A magnetically driven equatorial jet in Europa’s
  ocean} {A magnetically driven equatorial jet in europa’s ocean}.{\BBCQ}
\newblock
\APACjournalVolNumPages{Nature Astronomy}{3}{5}{401}.
\PrintBackRefs{\CurrentBib}

\bibitem [\protect \citeauthoryear {%
Goldreich%
\ \BBA {} Mitchell%
}{%
Goldreich%
\ \BBA {} Mitchell%
}{%
{\protect \APACyear {2010}}%
}]{%
Goldreich-Mitchell-2010:elastic}
\APACinsertmetastar {%
Goldreich-Mitchell-2010:elastic}%
\begin{APACrefauthors}%
Goldreich, P\BPBI M.%
\BCBT {}\ \BBA {} Mitchell, J\BPBI L.%
\end{APACrefauthors}%
\unskip\
\newblock
\APACrefYearMonthDay{2010}{}{}.
\newblock
{\BBOQ}\APACrefatitle {Elastic ice shells of synchronous moons: {Implications}
  for cracks on {Europa} and non-synchronous rotation of {Titan}} {Elastic ice
  shells of synchronous moons: {Implications} for cracks on {Europa} and
  non-synchronous rotation of {Titan}}.{\BBCQ}
\newblock
\APACjournalVolNumPages{Icarus}{209}{2}{631--638}.
\PrintBackRefs{\CurrentBib}

\bibitem [\protect \citeauthoryear {%
Goldsby%
\ \BBA {} Kohlstedt%
}{%
Goldsby%
\ \BBA {} Kohlstedt%
}{%
{\protect \APACyear {2001}}%
}]{%
Goldsby-Kohlstedt-2001:superplastic}
\APACinsertmetastar {%
Goldsby-Kohlstedt-2001:superplastic}%
\begin{APACrefauthors}%
Goldsby, D.%
\BCBT {}\ \BBA {} Kohlstedt, D.%
\end{APACrefauthors}%
\unskip\
\newblock
\APACrefYearMonthDay{2001}{}{}.
\newblock
{\BBOQ}\APACrefatitle {Superplastic deformation of ice: Experimental
  observations} {Superplastic deformation of ice: Experimental
  observations}.{\BBCQ}
\newblock
\APACjournalVolNumPages{Journal of Geophysical
  Research}{106}{B6}{11017--11030}.
\PrintBackRefs{\CurrentBib}

\bibitem [\protect \citeauthoryear {%
Goodman%
}{%
Goodman%
}{%
{\protect \APACyear {2012}}%
}]{%
Goodman-2012:tilted}
\APACinsertmetastar {%
Goodman-2012:tilted}%
\begin{APACrefauthors}%
Goodman, J\BPBI C.%
\end{APACrefauthors}%
\unskip\
\newblock
\APACrefYearMonthDay{2012}{}{}.
\newblock
{\BBOQ}\APACrefatitle {Tilted geostrophic convection in icy world oceans caused
  by the horizontal component of the planetary rotation vector} {Tilted
  geostrophic convection in icy world oceans caused by the horizontal component
  of the planetary rotation vector}.{\BBCQ}
\newblock
\BIn{} \APACrefbtitle {American Geophysical Union, Fall Meeting 2012, abstract}
  {American geophysical union, fall meeting 2012, abstract}\ (\BPG~P51A-2017).
\PrintBackRefs{\CurrentBib}

\bibitem [\protect \citeauthoryear {%
Goodman%
, Collins%
, Marshall%
\BCBL {}\ \BBA {} Pierrehumbert%
}{%
Goodman%
\ \protect \BOthers {.}}{%
{\protect \APACyear {2004}}%
}]{%
Goodman-Collins-Marshall-et-al-2004:hydrothermal}
\APACinsertmetastar {%
Goodman-Collins-Marshall-et-al-2004:hydrothermal}%
\begin{APACrefauthors}%
Goodman, J\BPBI C.%
, Collins, G\BPBI C.%
, Marshall, J.%
\BCBL {}\ \BBA {} Pierrehumbert, R\BPBI T.%
\end{APACrefauthors}%
\unskip\
\newblock
\APACrefYearMonthDay{2004}{}{}.
\newblock
{\BBOQ}\APACrefatitle {Hydrothermal plume dynamics on {Europa}: {Implications}
  for chaos formation} {Hydrothermal plume dynamics on {Europa}: {Implications}
  for chaos formation}.{\BBCQ}
\newblock
\APACjournalVolNumPages{J. Geophys. Res.}{109}{E3}{}.
\PrintBackRefs{\CurrentBib}

\bibitem [\protect \citeauthoryear {%
Goodman%
\ \BBA {} Lenferink%
}{%
Goodman%
\ \BBA {} Lenferink%
}{%
{\protect \APACyear {2012}}%
}]{%
Goodman-Lenferink-2012:numerical}
\APACinsertmetastar {%
Goodman-Lenferink-2012:numerical}%
\begin{APACrefauthors}%
Goodman, J\BPBI C.%
\BCBT {}\ \BBA {} Lenferink, E.%
\end{APACrefauthors}%
\unskip\
\newblock
\APACrefYearMonthDay{2012}{}{}.
\newblock
{\BBOQ}\APACrefatitle {Numerical simulations of marine hydrothermal plumes for
  {Europa} and other icy worlds} {Numerical simulations of marine hydrothermal
  plumes for {Europa} and other icy worlds}.{\BBCQ}
\newblock
\APACjournalVolNumPages{Icarus}{221}{2}{970--983}.
\PrintBackRefs{\CurrentBib}

\bibitem [\protect \citeauthoryear {%
Greenberg%
\ \protect \BOthers {.}}{%
Greenberg%
\ \protect \BOthers {.}}{%
{\protect \APACyear {1998}}%
}]{%
Greenberg-Geissler-Hoppa-et-al-1998:tectonic}
\APACinsertmetastar {%
Greenberg-Geissler-Hoppa-et-al-1998:tectonic}%
\begin{APACrefauthors}%
Greenberg, R.%
, Geissler, P.%
, Hoppa, G.%
, Tufts, B\BPBI R.%
, Durda, D\BPBI D.%
, Pappalardo, R.%
\BDBL {}Carr, M\BPBI H.%
\end{APACrefauthors}%
\unskip\
\newblock
\APACrefYearMonthDay{1998}{}{}.
\newblock
{\BBOQ}\APACrefatitle {Tectonic processes on {Europa}: {Tidal} stresses,
  mechanical response, and visible features} {Tectonic processes on {Europa}:
  {Tidal} stresses, mechanical response, and visible features}.{\BBCQ}
\newblock
\APACjournalVolNumPages{Icarus}{135}{1}{64--78}.
\PrintBackRefs{\CurrentBib}

\bibitem [\protect \citeauthoryear {%
Greenberg%
\ \BBA {} Weidenschilling%
}{%
Greenberg%
\ \BBA {} Weidenschilling%
}{%
{\protect \APACyear {1984}}%
}]{%
Greenberg-Weidenschilling-1984:how}
\APACinsertmetastar {%
Greenberg-Weidenschilling-1984:how}%
\begin{APACrefauthors}%
Greenberg, R.%
\BCBT {}\ \BBA {} Weidenschilling, S\BPBI J.%
\end{APACrefauthors}%
\unskip\
\newblock
\APACrefYearMonthDay{1984}{}{}.
\newblock
{\BBOQ}\APACrefatitle {How fast do {Galilean} satellites spin?} {How fast do
  {Galilean} satellites spin?}{\BBCQ}
\newblock
\APACjournalVolNumPages{Icarus}{58}{2}{186--196}.
\PrintBackRefs{\CurrentBib}

\bibitem [\protect \citeauthoryear {%
Hand%
, Chyba%
, Priscu%
, Carlson%
\BCBL {}\ \BBA {} Nealson%
}{%
Hand%
\ \protect \BOthers {.}}{%
{\protect \APACyear {2009}}%
}]{%
Hand-Chyba-Priscu-et-al-2009:astrobiology}
\APACinsertmetastar {%
Hand-Chyba-Priscu-et-al-2009:astrobiology}%
\begin{APACrefauthors}%
Hand, K.%
, Chyba, C.%
, Priscu, J.%
, Carlson, R.%
\BCBL {}\ \BBA {} Nealson, K.%
\end{APACrefauthors}%
\unskip\
\newblock
\APACrefYearMonthDay{2009}{}{}.
\newblock
{\BBOQ}\APACrefatitle {Astrobiology and the potential for life on {Europa}}
  {Astrobiology and the potential for life on {Europa}}.{\BBCQ}
\newblock
\APACjournalVolNumPages{Europa. University of Arizona Press,
  Tucson}{}{}{589--629}.
\PrintBackRefs{\CurrentBib}

\bibitem [\protect \citeauthoryear {%
Helfenstein%
\ \BBA {} Parmentier%
}{%
Helfenstein%
\ \BBA {} Parmentier%
}{%
{\protect \APACyear {1985}}%
}]{%
Helfenstein-Parmentier-1985:patterns}
\APACinsertmetastar {%
Helfenstein-Parmentier-1985:patterns}%
\begin{APACrefauthors}%
Helfenstein, P.%
\BCBT {}\ \BBA {} Parmentier, E.%
\end{APACrefauthors}%
\unskip\
\newblock
\APACrefYearMonthDay{1985}{}{}.
\newblock
{\BBOQ}\APACrefatitle {Patterns of fracture and tidal stresses due to
  nonsynchronous rotation: Implications for fracturing on {Europa}} {Patterns
  of fracture and tidal stresses due to nonsynchronous rotation: Implications
  for fracturing on {Europa}}.{\BBCQ}
\newblock
\APACjournalVolNumPages{Icarus}{61}{2}{175--184}.
\PrintBackRefs{\CurrentBib}

\bibitem [\protect \citeauthoryear {%
Hoppa%
\ \protect \BOthers {.}}{%
Hoppa%
\ \protect \BOthers {.}}{%
{\protect \APACyear {1999}}%
}]{%
Hoppa-Greenberg-Geissler-et-al-1999:rotation}
\APACinsertmetastar {%
Hoppa-Greenberg-Geissler-et-al-1999:rotation}%
\begin{APACrefauthors}%
Hoppa, G.%
, Greenberg, R.%
, Geissler, P.%
, Tufts, B\BPBI R.%
, Plassmann, J.%
\BCBL {}\ \BBA {} Durda, D\BPBI D.%
\end{APACrefauthors}%
\unskip\
\newblock
\APACrefYearMonthDay{1999}{}{}.
\newblock
{\BBOQ}\APACrefatitle {Rotation of {Europa}: Constraints from terminator and
  limb positions} {Rotation of {Europa}: Constraints from terminator and limb
  positions}.{\BBCQ}
\newblock
\APACjournalVolNumPages{Icarus}{137}{2}{341--347}.
\PrintBackRefs{\CurrentBib}

\bibitem [\protect \citeauthoryear {%
Hussmann%
, Spohn%
\BCBL {}\ \BBA {} Wieczerkowski%
}{%
Hussmann%
\ \protect \BOthers {.}}{%
{\protect \APACyear {2002}}%
}]{%
Hussmann-Spohn-Wieczerkowski-2002:thermal}
\APACinsertmetastar {%
Hussmann-Spohn-Wieczerkowski-2002:thermal}%
\begin{APACrefauthors}%
Hussmann, H.%
, Spohn, T.%
\BCBL {}\ \BBA {} Wieczerkowski, K.%
\end{APACrefauthors}%
\unskip\
\newblock
\APACrefYearMonthDay{2002}{}{}.
\newblock
{\BBOQ}\APACrefatitle {Thermal Equilibrium States of {Europa}'s Ice Shell:
  Implications for Internal Ocean Thickness and Surface Heat Flow} {Thermal
  equilibrium states of {Europa}'s ice shell: Implications for internal ocean
  thickness and surface heat flow}.{\BBCQ}
\newblock
\APACjournalVolNumPages{Icarus}{156}{}{143-151}.
\PrintBackRefs{\CurrentBib}

\bibitem [\protect \citeauthoryear {%
Kang%
}{%
Kang%
}{%
{\protect \APACyear {2022}}%
}]{%
Kang-2022:different}
\APACinsertmetastar {%
Kang-2022:different}%
\begin{APACrefauthors}%
Kang, W.%
\end{APACrefauthors}%
\unskip\
\newblock
\APACrefYearMonthDay{2022}{}{}.
\newblock
{\BBOQ}\APACrefatitle {Different Ice-shell Geometries on {Europa} and
  {Enceladus} due to Their Different Sizes: Impacts of Ocean Heat Transport}
  {Different ice-shell geometries on {Europa} and {Enceladus} due to their
  different sizes: Impacts of ocean heat transport}.{\BBCQ}
\newblock
\APACjournalVolNumPages{The Astrophysical Journal}{934}{2}{116}.
\PrintBackRefs{\CurrentBib}

\bibitem [\protect \citeauthoryear {%
Kang%
, Mittal%
, Bire%
, Campin%
\BCBL {}\ \BBA {} Marshall%
}{%
Kang%
\ \protect \BOthers {.}}{%
{\protect \APACyear {2022}}%
}]{%
Kang-Mittal-Bire-et-al-2022:how}
\APACinsertmetastar {%
Kang-Mittal-Bire-et-al-2022:how}%
\begin{APACrefauthors}%
Kang, W.%
, Mittal, T.%
, Bire, S.%
, Campin, J\BHBI M.%
\BCBL {}\ \BBA {} Marshall, J.%
\end{APACrefauthors}%
\unskip\
\newblock
\APACrefYearMonthDay{2022}{}{}.
\newblock
{\BBOQ}\APACrefatitle {How does salinity shape ocean circulation and ice
  geometry on {Enceladus} and other icy satellites?} {How does salinity shape
  ocean circulation and ice geometry on {Enceladus} and other icy
  satellites?}{\BBCQ}
\newblock
\APACjournalVolNumPages{Science Advances}{8}{29}{eabm4665}.
\PrintBackRefs{\CurrentBib}

\bibitem [\protect \citeauthoryear {%
Khurana%
\ \protect \BOthers {.}}{%
Khurana%
\ \protect \BOthers {.}}{%
{\protect \APACyear {1998}}%
}]{%
Khurana-Kivelson-Stevenson-et-al-1998:induced}
\APACinsertmetastar {%
Khurana-Kivelson-Stevenson-et-al-1998:induced}%
\begin{APACrefauthors}%
Khurana, K.%
, Kivelson, M.%
, Stevenson, D.%
, Schubert, G.%
, Russell, C.%
, Walker, R.%
\BCBL {}\ \BBA {} Polanskey, C.%
\end{APACrefauthors}%
\unskip\
\newblock
\APACrefYearMonthDay{1998}{}{}.
\newblock
{\BBOQ}\APACrefatitle {Induced magnetic fields as evidence for subsurface
  oceans in {Europa} and {Callisto}} {Induced magnetic fields as evidence for
  subsurface oceans in {Europa} and {Callisto}}.{\BBCQ}
\newblock
\APACjournalVolNumPages{Nature}{395}{6704}{777--780}.
\PrintBackRefs{\CurrentBib}

\bibitem [\protect \citeauthoryear {%
Kivelson%
\ \protect \BOthers {.}}{%
Kivelson%
\ \protect \BOthers {.}}{%
{\protect \APACyear {2000}}%
}]{%
Kivelson-Khurana-Russell-et-al-2000:galileo}
\APACinsertmetastar {%
Kivelson-Khurana-Russell-et-al-2000:galileo}%
\begin{APACrefauthors}%
Kivelson, M\BPBI G.%
, Khurana, K\BPBI K.%
, Russell, C\BPBI T.%
, Volwerk, M.%
, Walker, R\BPBI J.%
\BCBL {}\ \BBA {} Zimmer, C.%
\end{APACrefauthors}%
\unskip\
\newblock
\APACrefYearMonthDay{2000}{}{}.
\newblock
{\BBOQ}\APACrefatitle {Galileo magnetometer measurements: A stronger case for a
  subsurface ocean at {Europa}} {Galileo magnetometer measurements: A stronger
  case for a subsurface ocean at {Europa}}.{\BBCQ}
\newblock
\APACjournalVolNumPages{Science}{289}{5483}{1340--1343}.
\PrintBackRefs{\CurrentBib}

\bibitem [\protect \citeauthoryear {%
Lemasquerier%
\ \protect \BOthers {.}}{%
Lemasquerier%
\ \protect \BOthers {.}}{%
{\protect \APACyear {2017}}%
}]{%
Lemasquerier-Grannan-Vidal-et-al-2017:libration}
\APACinsertmetastar {%
Lemasquerier-Grannan-Vidal-et-al-2017:libration}%
\begin{APACrefauthors}%
Lemasquerier, D.%
, Grannan, A.%
, Vidal, J.%
, C{\'e}bron, D.%
, Favier, B.%
, Le~Bars, M.%
\BCBL {}\ \BBA {} Aurnou, J.%
\end{APACrefauthors}%
\unskip\
\newblock
\APACrefYearMonthDay{2017}{}{}.
\newblock
{\BBOQ}\APACrefatitle {Libration-driven flows in ellipsoidal shells}
  {Libration-driven flows in ellipsoidal shells}.{\BBCQ}
\newblock
\APACjournalVolNumPages{Journal of Geophysical Research:
  Planets}{122}{9}{1926--1950}.
\PrintBackRefs{\CurrentBib}

\bibitem [\protect \citeauthoryear {%
Losch%
}{%
Losch%
}{%
{\protect \APACyear {2008}}%
}]{%
Losch-2008:modeling}
\APACinsertmetastar {%
Losch-2008:modeling}%
\begin{APACrefauthors}%
Losch, M.%
\end{APACrefauthors}%
\unskip\
\newblock
\APACrefYearMonthDay{2008}{}{}.
\newblock
{\BBOQ}\APACrefatitle {Modeling ice shelf cavities in a z-coordinate ocean
  general circulation model} {Modeling ice shelf cavities in a z-coordinate
  ocean general circulation model}.{\BBCQ}
\newblock
\APACjournalVolNumPages{J. Geophys. Res.}{113}{}{C08043}.
\PrintBackRefs{\CurrentBib}

\bibitem [\protect \citeauthoryear {%
Marshall%
, Adcroft%
, Hill%
, Perelman%
\BCBL {}\ \BBA {} Heisey%
}{%
Marshall%
\ \protect \BOthers {.}}{%
{\protect \APACyear {1997}}%
}]{%
Marshall-Adcroft-Hill-et-al-1997:finite}
\APACinsertmetastar {%
Marshall-Adcroft-Hill-et-al-1997:finite}%
\begin{APACrefauthors}%
Marshall, J.%
, Adcroft, A.%
, Hill, C.%
, Perelman, L.%
\BCBL {}\ \BBA {} Heisey, C.%
\end{APACrefauthors}%
\unskip\
\newblock
\APACrefYearMonthDay{1997}{}{}.
\newblock
{\BBOQ}\APACrefatitle {A finite-volume, incompressible {Navier} {Stokes} model
  for studies of the ocean on parallel computers} {A finite-volume,
  incompressible {Navier} {Stokes} model for studies of the ocean on parallel
  computers}.{\BBCQ}
\newblock
\APACjournalVolNumPages{J. Geophys. Res.}{102, C3}{}{5,753--5,766}.
\PrintBackRefs{\CurrentBib}

\bibitem [\protect \citeauthoryear {%
McEwen%
}{%
McEwen%
}{%
{\protect \APACyear {1986}}%
}]{%
McEwen-1986:tidal}
\APACinsertmetastar {%
McEwen-1986:tidal}%
\begin{APACrefauthors}%
McEwen, A\BPBI S.%
\end{APACrefauthors}%
\unskip\
\newblock
\APACrefYearMonthDay{1986}{}{}.
\newblock
{\BBOQ}\APACrefatitle {Tidal reorientation and the fracturing of {Jupiter}'s
  moon Europa} {Tidal reorientation and the fracturing of {Jupiter}'s moon
  europa}.{\BBCQ}
\newblock
\APACjournalVolNumPages{Nature}{321}{6065}{49--51}.
\PrintBackRefs{\CurrentBib}

\bibitem [\protect \citeauthoryear {%
McKinnon%
}{%
McKinnon%
}{%
{\protect \APACyear {1999}}%
}]{%
McKinnon-1999:convective}
\APACinsertmetastar {%
McKinnon-1999:convective}%
\begin{APACrefauthors}%
McKinnon, W\BPBI B.%
\end{APACrefauthors}%
\unskip\
\newblock
\APACrefYearMonthDay{1999}{}{}.
\newblock
{\BBOQ}\APACrefatitle {Convective instability in {Europa}'s floating ice shell}
  {Convective instability in {Europa}'s floating ice shell}.{\BBCQ}
\newblock
\APACjournalVolNumPages{Geophysical Research Letters}{26}{7}{951--954}.
\PrintBackRefs{\CurrentBib}

\bibitem [\protect \citeauthoryear {%
Melosh%
, Ekholm%
, Showman%
\BCBL {}\ \BBA {} Lorenz%
}{%
Melosh%
\ \protect \BOthers {.}}{%
{\protect \APACyear {2004}}%
}]{%
Melosh-Ekholm-Showman-et-al-2004:temperature}
\APACinsertmetastar {%
Melosh-Ekholm-Showman-et-al-2004:temperature}%
\begin{APACrefauthors}%
Melosh, H.%
, Ekholm, A.%
, Showman, A.%
\BCBL {}\ \BBA {} Lorenz, R.%
\end{APACrefauthors}%
\unskip\
\newblock
\APACrefYearMonthDay{2004}{}{}.
\newblock
{\BBOQ}\APACrefatitle {The temperature of {Europa's} subsurface water ocean}
  {The temperature of {Europa's} subsurface water ocean}.{\BBCQ}
\newblock
\APACjournalVolNumPages{Icarus}{168}{2}{498--502}.
\PrintBackRefs{\CurrentBib}

\bibitem [\protect \citeauthoryear {%
{MITgcm~Group}%
}{%
{MITgcm~Group}%
}{%
{\protect \APACyear {2021}}%
}]{%
MITgcm-manual-github:mitgcm}
\APACinsertmetastar {%
MITgcm-manual-github:mitgcm}%
\begin{APACrefauthors}%
{MITgcm~Group}.%
\end{APACrefauthors}%
\unskip\
\newblock
\APACrefYearMonthDay{2021}{}{}.
\newblock
\APACrefbtitle {{MITgcm} {U}ser {M}anual} {{MITgcm} {U}ser {M}anual}\
  \APACbVolEdTR {}{Online documentation}.
\newblock
\APACaddressInstitution{Cambridge, MA 02139, USA}{{MIT}/{EAPS}}.
\newblock
\APACrefnote{\rm{https://mitgcm.readthedocs.io/en/latest/}}
\PrintBackRefs{\CurrentBib}

\bibitem [\protect \citeauthoryear {%
Moore%
\ \BBA {} Schubert%
}{%
Moore%
\ \BBA {} Schubert%
}{%
{\protect \APACyear {2000}}%
}]{%
Moore-Schubert-2000:tidal}
\APACinsertmetastar {%
Moore-Schubert-2000:tidal}%
\begin{APACrefauthors}%
Moore, W\BPBI B.%
\BCBT {}\ \BBA {} Schubert, G.%
\end{APACrefauthors}%
\unskip\
\newblock
\APACrefYearMonthDay{2000}{}{}.
\newblock
{\BBOQ}\APACrefatitle {The tidal response of {Europa}} {The tidal response of
  {Europa}}.{\BBCQ}
\newblock
\APACjournalVolNumPages{Icarus}{147}{1}{317--319}.
\PrintBackRefs{\CurrentBib}

\bibitem [\protect \citeauthoryear {%
Morrison%
}{%
Morrison%
}{%
{\protect \APACyear {2001}}%
}]{%
Morrison-2001:understanding}
\APACinsertmetastar {%
Morrison-2001:understanding}%
\begin{APACrefauthors}%
Morrison, F\BPBI A.%
\end{APACrefauthors}%
\unskip\
\newblock
\APACrefYear{2001}.
\newblock
\APACrefbtitle {Understanding rheology} {Understanding rheology}\ (\BVOL~1).
\newblock
\APACaddressPublisher{}{Oxford university press New York}.
\PrintBackRefs{\CurrentBib}

\bibitem [\protect \citeauthoryear {%
Ojakangas%
\ \BBA {} Stevenson%
}{%
Ojakangas%
\ \BBA {} Stevenson%
}{%
{\protect \APACyear {1989}}%
}]{%
Ojakangas-Stevenson-1989:thermal}
\APACinsertmetastar {%
Ojakangas-Stevenson-1989:thermal}%
\begin{APACrefauthors}%
Ojakangas, G\BPBI W.%
\BCBT {}\ \BBA {} Stevenson, D\BPBI J.%
\end{APACrefauthors}%
\unskip\
\newblock
\APACrefYearMonthDay{1989}{}{}.
\newblock
{\BBOQ}\APACrefatitle {Thermal state of an ice shell on {Europa}} {Thermal
  state of an ice shell on {Europa}}.{\BBCQ}
\newblock
\APACjournalVolNumPages{Icarus}{81}{2}{220--241}.
\PrintBackRefs{\CurrentBib}

\bibitem [\protect \citeauthoryear {%
Pappalardo%
\ \protect \BOthers {.}}{%
Pappalardo%
\ \protect \BOthers {.}}{%
{\protect \APACyear {1999}}%
}]{%
Pappalardo-Belton-Breneman-et-al-1999:does}
\APACinsertmetastar {%
Pappalardo-Belton-Breneman-et-al-1999:does}%
\begin{APACrefauthors}%
Pappalardo, R.%
, Belton, M.%
, Breneman, H.%
, Carr, M.%
, Chapman, C.%
, Collins, G.%
\BDBL {}others%
\end{APACrefauthors}%
\unskip\
\newblock
\APACrefYearMonthDay{1999}{}{}.
\newblock
{\BBOQ}\APACrefatitle {Does {Europa} have a subsurface ocean? {Evaluation} of
  the geological evidence} {Does {Europa} have a subsurface ocean? {Evaluation}
  of the geological evidence}.{\BBCQ}
\newblock
\APACjournalVolNumPages{J. Geophys. Res.}{104}{E10}{24015--24055}.
\PrintBackRefs{\CurrentBib}

\bibitem [\protect \citeauthoryear {%
Pappalardo%
\ \protect \BOthers {.}}{%
Pappalardo%
\ \protect \BOthers {.}}{%
{\protect \APACyear {2013}}%
}]{%
Pappalardo-Vance-Bagenal-et-al-2013:science}
\APACinsertmetastar {%
Pappalardo-Vance-Bagenal-et-al-2013:science}%
\begin{APACrefauthors}%
Pappalardo, R.%
, Vance, S.%
, Bagenal, F.%
, Bills, B.%
, Blaney, D.%
, Blankenship, D.%
\BDBL {}others%
\end{APACrefauthors}%
\unskip\
\newblock
\APACrefYearMonthDay{2013}{}{}.
\newblock
{\BBOQ}\APACrefatitle {Science potential from a {Europa} lander} {Science
  potential from a {Europa} lander}.{\BBCQ}
\newblock
\APACjournalVolNumPages{Astrobiology}{13}{8}{740--773}.
\PrintBackRefs{\CurrentBib}

\bibitem [\protect \citeauthoryear {%
{Rambaux}%
, {van Hoolst}%
\BCBL {}\ \BBA {} {Karatekin}%
}{%
{Rambaux}%
\ \protect \BOthers {.}}{%
{\protect \APACyear {2011}}%
}]{%
Rambaux-Hoolst-Karatekin-2011:librational}
\APACinsertmetastar {%
Rambaux-Hoolst-Karatekin-2011:librational}%
\begin{APACrefauthors}%
{Rambaux}, N.%
, {van Hoolst}, T.%
\BCBL {}\ \BBA {} {Karatekin}, {\"O}.%
\end{APACrefauthors}%
\unskip\
\newblock
\APACrefYearMonthDay{2011}{{\APACmonth{03}}}{}.
\newblock
{\BBOQ}\APACrefatitle {{Librational response of Europa, Ganymede, and Callisto
  with an ocean for a non-Keplerian orbit}} {{Librational response of Europa,
  Ganymede, and Callisto with an ocean for a non-Keplerian orbit}}.{\BBCQ}
\newblock
\APACjournalVolNumPages{Astron. Astrophys.}{527}{}{A118}.
\newblock
\begin{APACrefDOI} \doi{10.1051/0004-6361/201015304} \end{APACrefDOI}
\PrintBackRefs{\CurrentBib}

\bibitem [\protect \citeauthoryear {%
Roth%
\ \protect \BOthers {.}}{%
Roth%
\ \protect \BOthers {.}}{%
{\protect \APACyear {2014}}%
}]{%
Roth-Saur-Retherford-et-al-2014:transient}
\APACinsertmetastar {%
Roth-Saur-Retherford-et-al-2014:transient}%
\begin{APACrefauthors}%
Roth, L.%
, Saur, J.%
, Retherford, K\BPBI D.%
, Strobel, D\BPBI F.%
, Feldman, P\BPBI D.%
, McGrath, M\BPBI A.%
\BCBL {}\ \BBA {} Nimmo, F.%
\end{APACrefauthors}%
\unskip\
\newblock
\APACrefYearMonthDay{2014}{}{}.
\newblock
{\BBOQ}\APACrefatitle {Transient water vapor at {Europa's} south pole}
  {Transient water vapor at {Europa's} south pole}.{\BBCQ}
\newblock
\APACjournalVolNumPages{Science}{343}{6167}{171--174}.
\PrintBackRefs{\CurrentBib}

\bibitem [\protect \citeauthoryear {%
Rovira-Navarro%
\ \protect \BOthers {.}}{%
Rovira-Navarro%
\ \protect \BOthers {.}}{%
{\protect \APACyear {2019}}%
}]{%
Rovira-Navarro-Rieutord-Gerkema-et-al-2019:do}
\APACinsertmetastar {%
Rovira-Navarro-Rieutord-Gerkema-et-al-2019:do}%
\begin{APACrefauthors}%
Rovira-Navarro, M.%
, Rieutord, M.%
, Gerkema, T.%
, Maas, L\BPBI R.%
, van~der Wal, W.%
\BCBL {}\ \BBA {} Vermeersen, B.%
\end{APACrefauthors}%
\unskip\
\newblock
\APACrefYearMonthDay{2019}{}{}.
\newblock
{\BBOQ}\APACrefatitle {Do tidally-generated inertial waves heat the subsurface
  oceans of {Europa} and {Enceladus}?} {Do tidally-generated inertial waves
  heat the subsurface oceans of {Europa} and {Enceladus}?}{\BBCQ}
\newblock
\APACjournalVolNumPages{Icarus}{321}{}{126--140}.
\PrintBackRefs{\CurrentBib}

\bibitem [\protect \citeauthoryear {%
{Sarid}%
, {Greenberg}%
, {Hoppa}%
, {Geissler}%
\BCBL {}\ \BBA {} {Preblich}%
}{%
{Sarid}%
\ \protect \BOthers {.}}{%
{\protect \APACyear {2004}}%
}]{%
Sarid-Greenberg-Hoppa-et-al-2004:crack}
\APACinsertmetastar {%
Sarid-Greenberg-Hoppa-et-al-2004:crack}%
\begin{APACrefauthors}%
{Sarid}, A\BPBI R.%
, {Greenberg}, R.%
, {Hoppa}, G\BPBI V.%
, {Geissler}, P.%
\BCBL {}\ \BBA {} {Preblich}, B.%
\end{APACrefauthors}%
\unskip\
\newblock
\APACrefYearMonthDay{2004}{{\APACmonth{03}}}{}.
\newblock
{\BBOQ}\APACrefatitle {{Crack azimuths on Europa: time sequence in the southern
  leading face}} {{Crack azimuths on Europa: time sequence in the southern
  leading face}}.{\BBCQ}
\newblock
\APACjournalVolNumPages{Icarus}{168}{1}{144-157}.
\newblock
\begin{APACrefDOI} \doi{10.1016/j.icarus.2003.11.021} \end{APACrefDOI}
\PrintBackRefs{\CurrentBib}

\bibitem [\protect \citeauthoryear {%
Shoemaker%
\ \BBA {} Wolfe%
}{%
Shoemaker%
\ \BBA {} Wolfe%
}{%
{\protect \APACyear {1982}}%
}]{%
Shoemaker-Wolfe-1982:cratering}
\APACinsertmetastar {%
Shoemaker-Wolfe-1982:cratering}%
\begin{APACrefauthors}%
Shoemaker, E\BPBI M.%
\BCBT {}\ \BBA {} Wolfe, R\BPBI F.%
\end{APACrefauthors}%
\unskip\
\newblock
\APACrefYearMonthDay{1982}{}{}.
\newblock
{\BBOQ}\APACrefatitle {Cratering time scales for the {Galilean} satellites}
  {Cratering time scales for the {Galilean} satellites}.{\BBCQ}
\newblock
\BIn{} D.~Morrison\ (\BED), \APACrefbtitle {Satellites of Jupiter} {Satellites
  of jupiter}\ (\BPGS\ 277--339).
\newblock
\APACaddressPublisher{Tucson}{Univ. of Arizona Press}.
\PrintBackRefs{\CurrentBib}

\bibitem [\protect \citeauthoryear {%
Soderlund%
}{%
Soderlund%
}{%
{\protect \APACyear {2019}}%
}]{%
Soderlund-2019:ocean}
\APACinsertmetastar {%
Soderlund-2019:ocean}%
\begin{APACrefauthors}%
Soderlund, K\BPBI M.%
\end{APACrefauthors}%
\unskip\
\newblock
\APACrefYearMonthDay{2019}{}{}.
\newblock
{\BBOQ}\APACrefatitle {Ocean dynamics of outer solar system satellites} {Ocean
  dynamics of outer solar system satellites}.{\BBCQ}
\newblock
\APACjournalVolNumPages{Geophys. Res. Lett.}{}{}{doi:10.1029/2018GL081880}.
\PrintBackRefs{\CurrentBib}

\bibitem [\protect \citeauthoryear {%
Soderlund%
, Schmidt%
, Wicht%
\BCBL {}\ \BBA {} Blankenship%
}{%
Soderlund%
\ \protect \BOthers {.}}{%
{\protect \APACyear {2014}}%
}]{%
Soderlund-Schmidt-Wicht-et-al-2014:ocean}
\APACinsertmetastar {%
Soderlund-Schmidt-Wicht-et-al-2014:ocean}%
\begin{APACrefauthors}%
Soderlund, K\BPBI M.%
, Schmidt, B\BPBI E.%
, Wicht, J.%
\BCBL {}\ \BBA {} Blankenship, D\BPBI D.%
\end{APACrefauthors}%
\unskip\
\newblock
\APACrefYearMonthDay{2014}{}{}.
\newblock
{\BBOQ}\APACrefatitle {Ocean-driven heating of {Europa's} icy shell at low
  latitudes} {Ocean-driven heating of {Europa's} icy shell at low
  latitudes}.{\BBCQ}
\newblock
\APACjournalVolNumPages{Nature Geoscience}{7}{}{16-19}.
\PrintBackRefs{\CurrentBib}

\bibitem [\protect \citeauthoryear {%
Sparks%
\ \protect \BOthers {.}}{%
Sparks%
\ \protect \BOthers {.}}{%
{\protect \APACyear {2016}}%
}]{%
Sparks-Hand-McGrath-et-al-2016:probing}
\APACinsertmetastar {%
Sparks-Hand-McGrath-et-al-2016:probing}%
\begin{APACrefauthors}%
Sparks, W.%
, Hand, K.%
, McGrath, M.%
, Bergeron, E.%
, Cracraft, M.%
\BCBL {}\ \BBA {} Deustua, S.%
\end{APACrefauthors}%
\unskip\
\newblock
\APACrefYearMonthDay{2016}{}{}.
\newblock
{\BBOQ}\APACrefatitle {Probing for evidence of plumes on {Europa with
  HST/STIS}} {Probing for evidence of plumes on {Europa with HST/STIS}}.{\BBCQ}
\newblock
\APACjournalVolNumPages{The Astrophysical Journal}{829}{2}{121}.
\PrintBackRefs{\CurrentBib}

\bibitem [\protect \citeauthoryear {%
Thomson%
\ \BBA {} Delaney%
}{%
Thomson%
\ \BBA {} Delaney%
}{%
{\protect \APACyear {2001}}%
}]{%
Thomson-Delaney-2001:evidence}
\APACinsertmetastar {%
Thomson-Delaney-2001:evidence}%
\begin{APACrefauthors}%
Thomson, R\BPBI E.%
\BCBT {}\ \BBA {} Delaney, J\BPBI R.%
\end{APACrefauthors}%
\unskip\
\newblock
\APACrefYearMonthDay{2001}{}{}.
\newblock
{\BBOQ}\APACrefatitle {Evidence for a weakly stratified {Europan} ocean
  sustained by seafloor heat flux} {Evidence for a weakly stratified {Europan}
  ocean sustained by seafloor heat flux}.{\BBCQ}
\newblock
\APACjournalVolNumPages{J. Geophys. Res.}{106}{E6}{12355--12365}.
\PrintBackRefs{\CurrentBib}

\bibitem [\protect \citeauthoryear {%
Tobie%
, Choblet%
\BCBL {}\ \BBA {} Sotin%
}{%
Tobie%
\ \protect \BOthers {.}}{%
{\protect \APACyear {2003}}%
}]{%
Tobie-Choblet-Sotin-2003:tidally}
\APACinsertmetastar {%
Tobie-Choblet-Sotin-2003:tidally}%
\begin{APACrefauthors}%
Tobie, G.%
, Choblet, G.%
\BCBL {}\ \BBA {} Sotin, C.%
\end{APACrefauthors}%
\unskip\
\newblock
\APACrefYearMonthDay{2003}{}{}.
\newblock
{\BBOQ}\APACrefatitle {Tidally heated convection: Constraints on {Europa}'s ice
  shell thickness} {Tidally heated convection: Constraints on {Europa}'s ice
  shell thickness}.{\BBCQ}
\newblock
\APACjournalVolNumPages{J. Geophys. Res.}{108}{E11}{5124}.
\PrintBackRefs{\CurrentBib}

\bibitem [\protect \citeauthoryear {%
Tyler%
}{%
Tyler%
}{%
{\protect \APACyear {2008}}%
}]{%
Tyler-2008:strong}
\APACinsertmetastar {%
Tyler-2008:strong}%
\begin{APACrefauthors}%
Tyler, R\BPBI H.%
\end{APACrefauthors}%
\unskip\
\newblock
\APACrefYearMonthDay{2008}{}{}.
\newblock
{\BBOQ}\APACrefatitle {Strong ocean tidal flow and heating on moons of the
  outer planets} {Strong ocean tidal flow and heating on moons of the outer
  planets}.{\BBCQ}
\newblock
\APACjournalVolNumPages{Nature}{456}{7223}{770--772}.
\PrintBackRefs{\CurrentBib}

\bibitem [\protect \citeauthoryear {%
Vance%
\ \BBA {} Goodman%
}{%
Vance%
\ \BBA {} Goodman%
}{%
{\protect \APACyear {2009}}%
}]{%
Vance-Goodman-2009:oceanography}
\APACinsertmetastar {%
Vance-Goodman-2009:oceanography}%
\begin{APACrefauthors}%
Vance, S.%
\BCBT {}\ \BBA {} Goodman, J.%
\end{APACrefauthors}%
\unskip\
\newblock
\APACrefYearMonthDay{2009}{}{}.
\newblock
{\BBOQ}\APACrefatitle {Oceanography of an ice-covered moon} {Oceanography of an
  ice-covered moon}.{\BBCQ}
\newblock
\BIn{} R\BPBI T.~Pappalardo, W\BPBI B.~McKinnon\BCBL {}\ \BBA {} K.~Khurana\
  (\BEDS), \APACrefbtitle {Europa} {Europa}\ (\BPG~459-482).
\newblock
\APACaddressPublisher{}{The University of Arizona Press, Tucson, AZ}.
\PrintBackRefs{\CurrentBib}

\bibitem [\protect \citeauthoryear {%
Zeng%
\ \BBA {} Jansen%
}{%
Zeng%
\ \BBA {} Jansen%
}{%
{\protect \APACyear {2021}}%
}]{%
Zeng-Jansen-2021:ocean}
\APACinsertmetastar {%
Zeng-Jansen-2021:ocean}%
\begin{APACrefauthors}%
Zeng, Y.%
\BCBT {}\ \BBA {} Jansen, M\BPBI F.%
\end{APACrefauthors}%
\unskip\
\newblock
\APACrefYearMonthDay{2021}{}{}.
\newblock
{\BBOQ}\APACrefatitle {Ocean Circulation on Enceladus With a High Versus Low
  Salinity Ocean} {Ocean circulation on enceladus with a high versus low
  salinity ocean}.{\BBCQ}
\newblock
\APACjournalVolNumPages{arXiv preprint arXiv:2101.10530}{}{}{}.
\PrintBackRefs{\CurrentBib}

\end{thebibliography}

\end{document}